\documentclass[a4paper,12pt]{article}
\usepackage{jheppub,esint,shuffle,psfrag}
 \usepackage[utf8]{inputenc}
 
\usepackage{tikz}
\usetikzlibrary{arrows.meta,decorations.markings}

\usepackage{esint} 
\usepackage{breqn}
\usepackage{bm}
\usepackage{caption}
\usepackage{amssymb}
\usepackage{placeins}\input

\def \Im {\mathop{\rm Im}\nolimits}
\def \Re {\mathop{\rm Re}\nolimits}

\newcommand\lr[1]{{\left({#1}\right)}}

\def \qqqquad {\qquad\qquad}

\usepackage{comment}

\def\numberbysection{\@addtoreset{equation}{section}
                     \def\theequation{\thesection.\arabic{equation}}}

\begin{document}

\vspace*{0cm }

\author{Bercel Boldis$^{a,b}$}
\affiliation{
$\null$
$^a$Department of Theoretical Physics, Institute of Physics, Budapest University of Technology and Economics M\H{u}egyetem rkp. 3., 1111 Budapest, Hungary
\\
$\null$ 
$^b$HUN-REN Wigner Research Centre for Physics, Konkoly-Thege Miklos ut 29-33, 1121 Budapest, Hungary
  
}
\title{Strong coupling structure of $\mathcal{N}=4$ SYM observables with matrix Bessel kernel
}
\abstract{%
\small
In this paper I continue the program of studying the strong coupling expansion of certain observables in $\mathcal{N}=4$ supersymmetric Yang--Mills theory, which are given by a determinant with a matrix Bessel kernel. I show that, by reorganizing the transseries of the determinant at large values of the 't~Hooft coupling, a simple underlying structure emerges, in which each exponentially suppressed correction is related to the perturbative series in a simple way. This new approach provides an efficient method to generate the full transseries for $\mathcal{N}=4$ SYM observables, such as the cusp anomalous dimension, multi-gluon scattering amplitudes, and the octagon form factor. Using high-precision numerical analysis, I verify the results and provide a complete description of the resurgence structure of the strong coupling expansion.
}

\maketitle

\section{Introduction and summary}

Computing observables in four-dimensional superconformal Yang--Mills theories at finite 't~Hooft coupling $\lambda=g_{\mathrm{YM}}^2 N$ is central to the study of gauge/gravity duality. Within the AdS/CFT correspondence \cite{Aharony:1999ti}, observables in the planar limit of the gauge theory admit different descriptions depending on the value of $\lambda$. At weak coupling, they can be computed directly in the gauge theory using standard perturbative techniques. In this case, the expansion in the coupling constant is well controlled, and calculations can be systematically organized in terms of Feynman diagrams. However, for larger values of $\lambda$, the series typically become divergent and the weak coupling expansion loses its validity.

In contrast, at strong ’t~Hooft coupling, the same observables are described by a dual string theory. In this regime, the leading contribution is typically captured by a classical or semiclassical string configuration, whereas subleading terms arise from quantum fluctuations of the string worldsheet. Determining these corrections from direct string theory calculations is extremely difficult; thus, obtaining a finite $\lambda$ answer for a given observable from the strong coupling expansion is a challenging task.

Since accessing observables at finite ’t~Hooft coupling lies beyond the reach of both perturbative gauge theory and semiclassical string theory, therefore, a complete description of observables requires alternative methods, such as those based on integrability or localization \cite{Aharony:1999ti,Beisert:2010jr,Pestun_2017}. In recent years, significant progress has been made in these directions to determine certain observables for arbitrary ’t~Hooft coupling. 

In this paper, I investigate a special class of these observables. In the planar limit, at arbitrary values of the coupling constant, they admit the representation as a determinant:
\begin{align}\label{F-def}
Z_\ell(g)   = \det\Big(\delta_{nm}+\mathbb K_{nm}(\alpha)\Big)\Big|_{1+\ell \le n,m<\infty}\,.
\end{align}
These observables depend on the effective coupling constant $g=\sqrt{\lambda}/(4\pi)$, which enters the observable via the elements of the semi-infinite matrix $\mathbb{K}(\alpha)$. This matrix is casted in the block form:
\begin{align}\label{K-alpha}
\mathbb K(\alpha)=2\cos\alpha \left[\begin{array}{rr}\cos\alpha\, K_{\text{oo}}  & \sin\alpha\,  K_{\text{oe}} \\[1.2mm] -\sin\alpha\,  K_{\text{eo}} & \cos\alpha\,  K_{\text{ee}} \end{array}\right]\,.
\end{align}
Here $\alpha$ is a real parameter. Each block in $\mathbb{K}(\alpha)$ is itself a semi-infinite matrix. Their entries are expressed as the integrals:
\begin{align}\label{K-ori}
K_{nm} =\sqrt{nm}\int_0^\infty  {dt\over t} \chi\left(\frac{\sqrt{t}}{2g}\right) J_n(\sqrt{t}) J_m(\sqrt{t})\,,
\end{align}
where $J_n(x)$ is the Bessel function. $K_{nm}$ is the so-called truncated Bessel kernel \cite{BasorEhrhardt03}. The subscripts of the blocks $K_{oo}$,$K_{eo}$, etc. in \eqref{K-alpha} denote the parity of the indices in $K_{n,m}$ that contribute to that specific block; namely, their elements are given by:
\begin{align}\notag
    [K_{oo}]_{n,m}&=K_{2n-1,2m-1}\,,&&[K_{oe}]_{n,m}=K_{2n-1,2m}\,,\\
    [K_{eo}]_{n,m}&=K_{2n,2m-1}\,,&&[K_{ee}]_{n,m}=K_{2n,2m}\,.
\end{align}
The parameter $\alpha$ can be considered as a mixing angle between the subspaces of $K_{nm}$ defined by integrating over even or odd Bessel functions. The coupling constant enters the determinant via the function $\chi(x)$, typically referred to as the symbol of the Bessel matrix.

Determinants similar to \eqref{F-def} have previously appeared in several contexts in the study of supersymmetric gauge theories \cite{Beisert:2006ez,Coronado:2018ypq,Coronado:2018cxj,Kostov:2019stn,Bargheer:2019kxb,Kostov:2019auq,Belitsky:2019fan,Bargheer:2019exp,Basso:2020xts,Belitsky:2020qrm,Belitsky:2020qir,Beccaria:2020hgy,Beccaria:2021vuc,Beccaria:2021hvt,Beccaria:2022ypy,Beccaria:2023kbl}. Hence the form \eqref{F-def} is quite generic. Beside the 't~Hooft coupling, $Z_{\ell}(g)$ depends on the two real parameters $\alpha$ and $\ell$, and the explicit form of the symbol $\chi(x)$. For different values of the parameters and different functions $\chi(x)$, the determinant describes various observables of supersymmetric gauge theories:

\begin{enumerate}
    \item For $\alpha=0$, the off-diagonal blocks in \eqref{K-alpha} vanish, and $Z(g)$ can be written as a product of the determinants of the diagonal blocks. For different forms of the symbol, and values of $\ell$, these determinants describe several observables in planar $\mathcal{N}=4$ SYM, for instance, the flux tube correlators \cite{Beisert:2006ez,Belitsky:2019fan,Basso:2020xts} and correlation functions of infinitely heavy half-BPS operators \cite{,Coronado:2018ypq,Coronado:2018cxj,Kostov:2019stn,Bargheer:2019kxb,Kostov:2019auq,Belitsky:2019fan,Bargheer:2019exp}. For other choices of the symbol, they also give the leading non-planar correction to the partition function of $\mathcal{N}=2$ SYM \cite{Beccaria:2020hgy,Beccaria:2021vuc,Beccaria:2021hvt,Beccaria:2022ypy,Beccaria:2023kbl} and appear in the study of correlation functions in $\mathcal{N}=2$ superconformal quiver gauge theories \cite{Billo:2021rdb,Korchemsky:2025eyc}. In another case, they coincide with the Tracy-Widom distribution that describes the eigenvalue distribution in the Laguerre ensemble near the hard edge \cite{Tracy:1993xj}.

    \item For $\alpha=\pi/4$ and with symbol:
    \begin{equation}\label{chi-orig}
        \chi(x)=\frac{2}{e^x-1}\,,
    \end{equation}
    the matrix \eqref{K-alpha} is known as the BES kernel \cite{Beisert:2006ez}, and the determinant \eqref{F-def}  governs the cusp anomalous dimension $\Gamma_{\mathrm{cusp}}$ of $\mathcal{N}=4$ SYM. $\Gamma_{\mathrm{cusp}}$ is given by the ratio of the determinants evaluated at $\ell=1$ and $\ell=0$ \cite{Beccaria:2022ypy,Bajnok20252}, namely:
    \begin{equation}\label{tilted-cusp}
        \Gamma_{\mathrm{cusp}}(g)=4g^2\frac{Z_{\ell=1}(g)}{Z_{\ell=0}(g)}\,.
    \end{equation}
    
    \item For the same symbol as in \eqref{chi-orig}, but for different values of $\alpha=\pi r$, with $r$ being a rational number, $Z_{\ell}(g)$ also appears in the study of multi-gluon scattering amplitudes and form factors in planar $\mathcal{N}=4$ SYM \cite{Basso:2020xts,Basso:2022ruw,Basso:2023bwv,Basso:2024hlx}.

\end{enumerate}

As discussed in \cite{Bajnok20252}, the determinant \eqref{F-def} can be reformulated in the form of a Fredholm determinant of an integral operator whose kernel can be written as a two-by-two matrix of Bessel kernels (for this reason, I simply refer to \eqref{F-def} as a determinant with a matrix Bessel kernel). In recent years a huge development was made in the study of Fredholm determinants with Bessel kernel. Based on previous results from the mathematical literature (see, for instance, \citep{Szego1915,Szego1952,Kac1954,Akhiezer1966,FisherHartwig1969,Bottcher1995OnsagerFisherHartwig,Basor2012BriefHistorySzego,Its1990DifferentialEF,Korepin:1993kvr,Tracy:1993xj}), it was shown that these determinants and the resolvents of the Bessel kernel satisfy a system of coupled integro-differential equations and they provide a systematic way to expand them both for weak and strong coupling in an efficient way \citep{Its1990,Belitsky:2020qrm,Belitsky:2020qir,Bajnok2024,Bajnok20251,Bajnok20252}. By high order numerical computations, it was shown that these series are asymptotic and a resurgence analysis \citep{Marino:2012zq,Dorigoni:2014hea,Aniceto:2018bis} is essential to obtain a finite physical answer. The resurgence properties for the weak and strong coupling regimes were studied in \cite{Basso:2009gh,Aniceto:2015rua, Dorigoni:2015dha,Bajnok2024,Bajnok20251,Bajnok20252,Dunne:2025wbq,bajnok20253}. Relations for the determinant \eqref{F-def} for different values of the parameter $\ell$ were also found in \cite{korchemsky20252}.

In this paper, I continue the program of \cite{Bajnok2024,Bajnok20251,Bajnok20252,bajnok20253} and investigate further properties of the strong-coupling expansion of \eqref{F-def} for arbitrary values of $\alpha$ and $\ell$, with the specific symbol \eqref{chi-orig}. In \cite{Bajnok20252} we have already presented a method to compute the large-$g$ expansion for $Z_{\ell}(g)$ in this case up to arbitrary orders in the transseries parameters and investigated its resurgence properties. Our method was based on first computing the logarithm of the determinant $\mathcal{F}_{\ell}(g)$ through a set of integro-differential equations and then exponentiating the resulting series to obtain the strong coupling expansion of $Z_{\ell}(g)=e^{\mathcal F_{\ell}(g)}$.

In this way, we found that at large $g$, the determinant $Z_\ell(\alpha)$ can be expanded as the transseries:
\begin{equation}\label{eq:trans_gamma}
    Z_\ell(g)=A_\ell(g)\sum_{n,m\geq 0}\Lambda_{-}^{2n}\Lambda_+^{2m}\mathcal{Z}^{(n,m)}(g)\,,
\end{equation}
where $a=\alpha/\pi$. Each coefficient function $\mathcal{Z}^{(n,m)}(g)$ is given by series in $1/g$. The exponentially small corrections -- which corresponds to $(n,m)\neq(0,0)$ -- are governed by the parameters 
\begin{align}\label{Lambda}  
 \Lambda_-^2 = g^{2a} e^{-4\pi g (1-2a)} \,,\qqqquad \Lambda_+^2= g^{-2a} e^{-4\pi g (1+2a)} \,.
\end{align}

The derivation of the large-$g$ expansion in \eqref{eq:trans_gamma} heavily relied on the analytical properties of the symbol in \eqref{chi-orig}, or more precisely, on the combination:
\begin{align}\label{Omega}
\chi_\alpha(x) =e^{i\alpha} +\chi(x) \cos\alpha={\cosh(x/2+i\alpha)\over \sinh (x/2)}\,.
\end{align}
The exponential scales in \eqref{Lambda} and their products defining the large-$g$ transseries are related to the zeros of this function, located away from the real axis at:
\begin{align}\label{zeros0}
    x=2\pi i\left(l+\frac{1}{2}- a\right)\equiv 2\pi i x_l^+,\qquad\text{and}\qquad x=-2\pi i\left(j+\frac{1}{2}+ a\right)\equiv -2\pi i x_j^-\,,
\end{align}
with $l,j\in\mathbb{N}_0$.

The prefactor $A(g)$ in \eqref{eq:trans_gamma} contains an overall dependence on $g$ and the Widom-Dyson constant $B_\ell(\alpha)$ \cite{Bajnok20252}:
\begin{equation}\label{B-guess}
    A_\ell(g)=\frac{e^{\pi(1-4a^2)g}}{(8\pi g)^{1/4+\ell+a^2}}e^{B_\ell(a)}\,.
\end{equation}
By numerically evaluating the determinant for different values of the angle $\alpha=\pi a$ and analytically investigating its behavior around $a=0$ and $a=1/2$, we previously found a fitting expression for Widom-Dyson constant for arbitrary $a$ and $\ell$ (see equation (5.18) of \cite{Bajnok20252}).

Although we were able to obtain several analytic expressions for the strong coupling coefficients of $Z_{\ell}(g)$, we observed cancellations of certain non-perturbative contributions, and non-trivial resurgence relations were found between the remaining corrections. The governing principle behind this structure remained unknown.

In our latest paper \cite{bajnok20253}, we have shown that in case of $\alpha=0$ (which we studied previously in \cite{Bajnok2024} and \cite{Bajnok20251}), there is a simple underlying structure in the large $g$ expansion of the determinant itself. We found that the non-perturbative contributions appear only at first orders in certain exponential scales, which are related to the zeros of the function $1-\chi(x)$. Furthermore, these contributions can be easily obtained from the perturbative sector by applying simple transformations on the coefficients to the perturbative $1/g$ series. We also found that in this new structure the resurgence between different non-perturbative sectors appears in a natural way.

The question then arises whether the same principle can be generalized to observables given by the matrix Bessel kernel defined in \eqref{K-alpha} with symbol \eqref{chi-orig} and arbitrary values of $\alpha$. In this paper, I demonstrate that, by rearranging the strong coupling expansion found in \cite{Bajnok20252}, the results of \cite{bajnok20253} can be extended to the observables discussed in points 2. and 3. above, making it extremely efficient to generate the complete transseries for these physical quantities as well. This new form of the transseries also provides the opportunity to give a complete description to the resurgence structure of the strong coupling expansion of these observables.

\subsection*{Summary and conclusions}

The paper is organized as follows: In Section 2, I summarize the results obtained in \cite{Bajnok20252} for the strong coupling expansion \eqref{eq:trans_gamma}. I show some examples of the transseries coefficients and I discuss that they have an explicit dependence on two different ingredients: the parameter $a=\alpha/\pi$ and a set of integrals $\mathcal I_n$ related to the moments of $\partial_x \log \chi_\alpha(x)$. I also motivate that there is an underlying structure that connects the exponentially suppressed corrections in the transseries to the perturbative sector.

In the first part of Section 3, I rearrange the transseries and show that the strong coupling expansion in \eqref{eq:trans_gamma} can be written in the form:
\begin{align}\label{trans-D-exp}
    Z_\ell(g)=A(g)&\sum_{\delta^+,\delta^-}(8\pi g)^{-\Delta(\Delta-2a)}e^{-8\pi g \left(\sum_{l\in \delta^+}  x_l^++\sum_{j\in \delta^- } x_j^-\right)}e^{i\pi a\Delta }S^{(\delta^+,\delta^-)}\mathcal{D}^{(\delta^+,\delta^-)}(g)\,,
\end{align}
The exponential weights $x_l^+$ and $x_j^-$ are the zeros of the function $\chi_\alpha (x)$ given in \eqref{zeros0}. The summation runs over all possible pairs of finite sets $\delta^+$ and $\delta^-$, both containing only non-negative integers. This ensures that all terms in the transseries are at most first order in each distinct exponential factor $e^{-8\pi gx_l^+}$ and $e^{-8\pi gx_j^-}$. For a fixed pair of $\delta^+$ and $\delta^-$, $\Delta$ denotes the difference between the number of elements in $\delta^+$ and $\delta^-$, that is, it is equal to $\Delta=|\delta^+|-|\delta^-|$. The functions $\mathcal{D}^{(\delta^+,\delta^-)}(g)$ are given by expansions over $1/ g$ and $S^{(\delta^+,\delta^-)}$ are the corresponding Stokes constants.

In Section 3.1, I show that it is convenient to introduce the notation for the perturbative $(\delta^+,\delta^-)=(\{\},\{\})$ sector:
\begin{equation}
    \mathcal{D}^{(\{\},\{\})}( g)=\mathcal{D}\left[\mathcal{I}_n\right](g)\,,
\end{equation}
to make the dependence on the moments $\mathcal I_n$ explicit. With the system of integro-differential equations studied in \cite{Bajnok20252}, the $1/g$ expansion of $\mathcal{D}\left[\mathcal{I}_n\right](g)$ can be efficiently generated both analytically and numerically. Then applying the same ideas as in \cite{bajnok20253}, I show that the exponentially small corrections can be easily generated from the perturbative functions with the simple rule:
\begin{equation}\label{D-rule-fin}
    \mathcal{D}^{(\delta^+,\delta^-)}(g)=\left.\mathcal{D}\left[\mathcal{I}^{(\delta^+,\delta^-)}_n\right](g)\right|_{a\to a-\Delta}\,.
\end{equation}
This expression means that the non-perturbative corrections $\mathcal{D}^{(\delta^+,\delta^-)}( g)$ are obtained from the perturbative sector by shifting the explicit $a$ dependence by $-\Delta$ and the moments should be replaced by suitably modified integrals denoted by $\mathcal{I}_n^{(\delta^+,\delta^-)}$ (for their explicit form see equation \eqref{I-mod-exact})

In Section 3.2, I use the same idea to show that there exist two recurrence relations, that generate all the corresponding Stokes constants $S^{(\delta^+,\delta^-)}$ as well. These relations are given in \eqref{Sp-fin} and \eqref{Sm-fin}. The rule \eqref{D-rule-fin} for the $1/g$-expansions together with the recurrence relations for the Stokes constants completely generates the full strong coupling expansion \eqref{trans-D-exp} of the determinant with matrix Bessel kernel in an effective way.

In Section 3.3, as a practical example, I specify the parameter $\alpha=\pi/4$. Using the relation between the exponentially small corrections and the perturbative sector, I present several analytic results for the strong coupling corrections for the determinant evaluated at the above-mentioned value of $\alpha$ and $\ell=0,1$. Then, by \eqref{tilted-cusp}, I also give some explicit non-perturbative corrections for the cusp anomalous dimension.

In Section 4, I investigate the resurgence properties of the transseries \eqref{trans-D-exp}. Through the Bridge equations (discussed in more detail in Appendix A) I derive the Alien algebra for \eqref{trans-D-exp} and I show that this transseries is more natural from the point of view of resurgence than the expansion given in \eqref{eq:trans_gamma}. With high precision numerical analysis of certain exponential corrections, I also explain some resurgence relations that were found in \cite{Bajnok20252}.

The advantage of writing the transseries as \eqref{trans-D} instead of \eqref{eq:trans_gamma}, is that the exponential scales are in one-to-one correspondence with the zeros of the relevant symbol \eqref{Omega}. Moreover, the exponentially small corrections are at most first order in every scale $e^{-8\pi g x_l^+}$ and $e^{-8\pi g x_j^-}$, and the corresponding $1/g$ expansions can be easily obtained from the perturbative sector with the simple rule in \eqref{D-rule-fin}. This is the same structure as was studied in \cite{bajnok20253}. Therefore, \eqref{trans-D} puts the observables in points 2. and 3. defined through the matrix Bessel kernel with different values of the mixing angle $\alpha$ in the universal strong coupling structure of \cite{bajnok20253}. It also provides a more natural framework for describing the resurgence properties of the above-mentioned observables at large values of the ’t Hooft coupling.

According to the results of this paper together with \cite{bajnok20253}, there is a simple underlying structure in the strong coupling expansion for certain observables of $\mathcal{N}=4$ SYM, at least for those given by the determinant \eqref{F-def}. It is an open question whether the same structure holds for other observables in $\mathcal{N}=4$ SYM as well, or physical quantities in different but related models, such as the $O(6)$ model \cite{Polyakov:1980ca,Korchemsky:1987wg,Alday:2007mf}. It would also be interesting to relate the obtained transseries structure to the results of \cite{Bajnok:2022xgx,Bajnok:2025mxi} where a complete resurgence analysis was carried out for a family of generalized energy densities in the $O(N)$ models.

\section{Strong coupling expansion}\label{sec:Strong-results}

In \cite{Bajnok20252} we presented a technique to determine the strong coupling expansion of the determinant \eqref{F-def} for arbitrary values of $a$ and $\ell$. Our method was based on a complicated system of integro-differential equations for the logarithm of the determinant $\mathcal{F}_\ell(g)=\log Z_\ell(g)$. With this, we were able to analytically compute the perturbative part $\mathcal{Z}^{(0,0)}(g)$ in \eqref{eq:trans_gamma} and many non-perturbative contributions $\mathcal{Z}^{(n,m)}$, with $(n,m)\neq (0,0)$ up to several orders in $1/g$. In the following, I highlight some of the analytic results for the perturbative and non-perturbative sectors of \eqref{eq:trans_gamma}.

\subsection*{Perturbative part}

First, I consider the perturbative sector given by $\mathcal{Z}^{(0,0)}(g)$. The $1/g$ expansion of $\mathcal{Z}^{(0,0)}(g)$ can  parametrized in the following way:
\begin{equation}\label{Z00}
    \mathcal{Z}^{(0,0)}(g)=1+(a^2-\ell^2)\sum_{k\geq 1}\frac{(-1)^k}{k!}\frac{f_k^{(0,0)}}{(8\pi g)^k}\,,
\end{equation}
with the first few functions $f_k^{(0,0)}$ being:
\begin{align}\notag\label{f00}
    f^{(0,0)}_1&=\mathcal{I}_2\,,\\
    f^{(0,0)}_2&=(a^2-\ell^2+1)\mathcal{I}_2^2+2a\mathcal{I}_3\,,\notag\\
    f^{(0,0)}_3&=\left(a^2-\ell ^2+1\right) \left(a^2-\ell ^2+2\right)\mathcal{I}_2^3+6 a \left(a^2-\ell ^2+2\right)\mathcal{I}_2\mathcal{I}_3+2(5 a^2- \ell ^2+1)\mathcal{I}_4\,,\notag\\
    f^{(0,0)}_4&=\left(a^2-\ell ^2+1\right) \left(a^2-\ell ^2+2\right) \left(a^2-\ell ^2+3\right)\mathcal{I}_2^4+12 a \left(a^2-\ell ^2+2\right) \left(a^2-\ell ^2+3\right)\mathcal{I}_2^2\mathcal{I}_3+\notag\\
    &+6 \left(2 a^4-2 a^2 \ell ^2+9 a^2-\ell ^2+1\right)\mathcal{I}_3^2+8 \left(a^2-\ell ^2+3\right) \left(5 a^2-\ell ^2+1\right)\mathcal{I}_2\mathcal{I}_4+\notag\\
    &+12 a \left(7 a^2-3 \ell ^2+5\right)\mathcal{I}_5\,.
\end{align}

The modified symbol \eqref{Omega} enters the perturbative coefficients $f_n^{(0,0)}$ through the moments\footnote{For convenience, compared to the notations of \cite{Bajnok20252}, I used a different normalization for the integrals of the symbol $\chi_\alpha(x)$. The integrals $I_n$ defined in \cite{Bajnok20252} are related to $\mathcal{I}_n$ via $\mathcal{I}_n=(2\pi)^{n-1}I_n$.}:
\begin{align}\label{In}
\mathcal{I}_n={1\over\pi^2}\Re\left[ \int_0^\infty {dz} \,\lr{2\pi i\over z}^n z \partial_z \log \chi_\alpha(z) \right],
\end{align}
Each function $f^{(0,0)}_n$ is a multilinear combination of $\mathcal{I}_n$. For $n\geq 2$, these integrals diverge at the origin; however, with analytical regularization, they can be evaluated, giving:
\begin{align}\label{In-def}
\mathcal{I}_{n}= {1\over (n-2)!}\bigg[ {}& \psi ^{(n-2)}\left(\frac{1}{2}-a\right)-\psi ^{(n-2)}(1)  +(-1)^n \bigg(\psi ^{(n-2)}\left(\frac{1}{2}+a\right)- \psi ^{(n-2)}(1)\bigg) \bigg]\,,
\end{align}
where $\psi^{(n)}(x)$ is the polygamma function of order $n$. Notice that the functions $\mathcal I_n$ have the symmetry under $a\to -a$:
\begin{equation}
    \mathcal{I}_n(a)=(-1)^n\mathcal{I}_n(-a)\,.
\end{equation}
The moments $\mathcal I_n$ depend on the mixing angle $a$. However, for reasons explained later, I suppress their $a$ dependence and indicate it explicitly only when necessary.

Following the method of \cite{Bajnok20252}, the perturbative part can be effectively determined up to arbitrary order in $1/g$.

\subsection*{Leading $(1,0)$ non-perturbative correction}

The determinant \eqref{F-def} is an even function of $\alpha$ and periodic in $\alpha\to \alpha+\pi$ therefore, it is enough to consider $0<\alpha<\pi/2$. In this case, for large values of $g$, $\Lambda_-\gg\Lambda_+$, which gives an ordering for the exponentially suppressed terms.  Hence the leading non-perturbative sector corresponds to $(n,m)=(1,0)$. Its $1/g$ expansion looks as:
\begin{align}\label{Z10}
    \mathcal{Z}^{(1,0)}(g)=e^{i\pi a}P^{(1,0)}\frac{(8\pi)^{2a}}{8\pi g}\left[1+((a-1)^2-\ell^2)\sum_{k\geq 1}\frac{(-1)^k}{k!}\frac{f_k^{(1,0)}}{(8\pi g)^k}\right]\,,
\end{align}
where the prefactor $(8\pi)^{2a}$ was introduced to match the power of $g$ included in $\Lambda_-^2$. The series starts at $\mathcal{O}(1/g)$, and the functions $f_n^{(1,0)}$ are again multilinear combinations of the integrals $\mathcal{I}_n$. Up to $1/g^3$ they are given by:
\begin{align}\notag\label{f10}
    f^{(1,0)}_1&=\mathcal{\mathcal{I}}_2+2(1/2-a)^{-1}\,,\notag\\
    f^{(1,0)}_2&=((a-1)^2-\ell^2+1)\left(\mathcal{I}_2+2(1/2-a)^{-1}\right)^2+2(a-1)\left(\mathcal{I}_3-2(1/2-a)^{-2}\right)\,,\notag\\
    f^{(1,0)}_3&=\left((a-1)^2-\ell ^2+1\right) \left((a-1)^2-\ell ^2+2\right)\left(\mathcal{I}_2+2(1/2-a)^{-1}\right)^3+\notag\\
    &+6 (a-1) \left((a-1)^2-\ell ^2+2\right)\left(\mathcal{I}_2+2(1/2-a)^{-1}\right)\left(\mathcal{I}_3-2(1/2-a)^{-2}\right)+\notag\\
    &+2(5 (a-1)^2- \ell ^2+1)\left(\mathcal{I}_4+2(1/2-a)^{-3}\right)\,,
\end{align}
and the prefactor $P^{(1,0)}$ is:
\begin{align}\label{P10}
    P^{(1,0)}=-(-1)^{\ell }\left(\frac{1}{2}-a\right)^{2 a-1} \frac{ \Gamma (1+\ell-a)}{ \Gamma (\ell+a )}\,.
\end{align}
It can be seen that the series $\mathcal{Z}^{(1,0)}$ is complex valued, as it contains an overall factor of $e^{i\pi a}$. The rest of the expansion is real.

Notice that functions $\mathcal{Z}^{(1,0)}$ and $f^{(1,0)}_n$ are related to $f^{(0,0)}_n$ corresponds to the perturbative part and is shown in \eqref{Z00} and \eqref{f00}. $f^{(1,0)}_n$, as well as their prefactors in the series \eqref{Z10} can be obtained from \eqref{Z00} and \eqref{f00} by simply shifting the explicit $a$ dependence by $a\to a-1$ and changing the integrals $\mathcal{I}_n$ as:
\begin{equation}\label{shift10}
    \mathcal{I}_2\to \mathcal{I}_2+\frac{2}{\left(\frac{1}{2}-a\right)}\,,\qquad \mathcal{I}_3\to \mathcal{I}_3-\frac{2}{\left(\frac{1}{2}-a\right)^2}\,,\qquad \mathcal{I}_4\to \mathcal{I}_4+\frac{2}{\left(\frac{1}{2}-a\right)^3}\,,\qquad\dots\,,
\end{equation}
or in general:
\begin{equation}\label{shift-gen-10}
    \mathcal{I}_n\to \mathcal{I}_n-(-1)^{n-1}\frac{2}{\left(\frac{1}{2}-a\right)^{n-1}}\,.
\end{equation}

\subsection*{$(0,1)$ correction}

The next subleading exponentially small contribution is $\Lambda_-^2\mathcal{Z}^{(0,1)}(g)$. Its $1/g$ expansion can be parametrized as:
\begin{align}\label{Z01}
    \mathcal{Z}^{(0,1)}(g)=P^{(0,1)}\frac{(8\pi)^{-2a}}{8\pi g}\left[1+((a+1)^2-\ell^2)\sum_{k\geq 1}\frac{(-1)^k}{k!}\frac{f_k^{(0,1)}}{(8\pi g)^k}\right]\,.
\end{align}
From the definition of $\mathbb{K}(\alpha)$ it follows that the determinant is symmetric under the exchange $\alpha\to -\alpha$, therefore the strong-coupling expansion should reflect this symmetry as well. This leads to a relation between the two non-perturbative sectors:
\begin{equation}
    \mathcal{Z}^{(0,1)}(g|a)=\mathcal{Z}^{(1,0)}(g|-a)\,,
\end{equation}
where the change $a\to -a$ applies to the full dependence on $a$. The same property for the corresponding exponential scales $\Lambda^2_-$ and $\Lambda^2_+$. Therefore the prefactor $P^{(0,1)}$ and the coefficients $f^{(0,1)}_k$ can be easily obtained from $P^{(1,0)}$ and $f^{(1,0)}_k$ by changing $a\to -a$.

The series $\mathcal{Z}^{(0,1)}$ is again complex valued, since it contains an overall factor of $e^{-i\pi a}$.

It is important to note, that up to an overall prefactor, the series \eqref{Z01} can be easily obtained also from the perturbative series \eqref{Z00}, by shifting the explicit $a$ dependence with $a\to a+1$ and transforming the integrals $\mathcal{I}_n$ as:
\begin{equation}\label{shift01}
    \mathcal{I}_2\to \mathcal{I}_2+\frac{2}{\left(\frac{1}{2}+a\right)}\,,\qquad \mathcal{I}_3\to \mathcal{I}_3+\frac{2}{\left(\frac{1}{2}+a\right)^2}\,,\qquad \mathcal{I}_4\to \mathcal{I}_4+\frac{2}{\left(\frac{1}{2}+a\right)^3}\,,\qquad\dots\,,
\end{equation}
or:
\begin{equation}\label{shift-gen-01}
    \mathcal I_n\to \mathcal I_n+\frac{2}{\left(\frac{1}{2}+a\right)^{n-1}}\,.
\end{equation}
The series again starts with a $\mathcal{O}(1/g)$ term.

\subsection*{$(1,1)$ correction}
The next subleading contribution appears at order $\Lambda_-^2\Lambda_+^2$. At this order we find the $1/g$ expansion:
\begin{equation}\label{Z11}
    \mathcal{Z}^{(1,1)}(g)=P^{(1,1)}\left[1+(a^2-\ell^2)\sum_{k\geq 1}\frac{(-1)^k}{k!}\frac{f_k^{(1,1)}}{(8\pi g)^k}\right]\,.
\end{equation}
The functions $f^{(1,1)}_n$ are similar to \eqref{f11}:
\begin{align}\notag\label{f11}
    f^{(1,1)}_1&=\mathcal{I}_2+2(1/2-a)^{-1}+2(1/2+a)^{-1}\,,\notag\\
    f^{(1,1)}_2&=(a^2-\ell^2+1)\left(\mathcal{I}_2+2(1/2-a)^{-1}+2(1/2+a)^{-1}\right)^2+\notag\\
    &+2a\left(\mathcal{I}_3-2(1/2-a)^{-2}+2(1/2+a)^{-2}\right)\,,\notag\\
    f^{(1,1)}_3&=\left(a^2-\ell ^2+1\right) \left(a^2-\ell ^2+2\right)\left(\mathcal{I}_2+2(1/2-a)^{-1}+2(1/2+a)^{-1}\right)^3+\notag\\
    &+6 a \left(a^2-\ell ^2+2\right)\left(\mathcal{I}_2+2(1/2-a)^{-1}+2(1/2+a)^{-1}\right)\times\notag\\
    &\times\left(\mathcal{I}_3-2(1/2-a)^{-2}+2(1/2+a)^{-2}\right)+\notag\\
    &+2(5 a^2- \ell ^2+1)\left(\mathcal{I}_4+2(1/2-a)^{-3}+2(1/2+a)^{-3}\right)\,.
\end{align}
The overall prefactor $P^{(1,1)}$ is given by:
\begin{equation}
    P^{(1,1)}=-\left(\frac{1}{2}-a\right)^{1+2 a} \left(\frac{1}{2}+a\right)^{1-2 a}\,.
\end{equation}
In this case $\mathcal{Z}^{(1,1)}(g)$ is purely real.

It can be easily seen, that $f_n^{(1,1)}$ can be directly obtained from the perturbative coefficients $f_n^{(0,0)}$ by first applying the shift \eqref{shift10} then \eqref{shift01} subsequently on the integrals $\mathcal{I}_n$, or vica versa.

\subsection*{General observations}

From the definition of $\mathbb{K}(\alpha)$ it follows that the determinant is symmetric under the exchange $\alpha\to -\alpha$, therefore the strong-coupling expansion should reflect this symmetry as well. This leads to relation between the non-perturbative sectors:
\begin{equation}\label{Z-symm}
    \mathcal Z^{(n,m)}(g|a)=\mathcal Z^{(m,n)}(g|-a)\,.
\end{equation}
where the change $a\to -a$ applies again to the full dependence on $a$. The exponential scales in \eqref{Lambda} also respect this symmetry.

By generating several large-$g$ contributions both analytically and numerically for the logarithm of the determinant $\mathcal{F}(g)$ and taking its exponential to compute the determinant, we observed the cancellation of several non-perturbative corrections in $Z_\ell(g)$. We found that $Z^{(n,m)}$ is only non-zero, if $(n,m)$ satisfies the following constraint:
\begin{align}\label{nonzero}
    \frac{1}{2}p(p-1)\leq n <\frac{1}{2}p(p+1)\,,\qquad 0\leq m-n \leq p\,,
\end{align}
with $p$ being an integer such that $p\geq 0$. In Figure 1, I show at which orders of $\Lambda_-^2$ and $\Lambda_+^2$ there is a non-zero series $\mathcal{Z}^{(n,m)}$.
\begin{figure}[t]
\begin{centering}
\includegraphics[width=9cm]{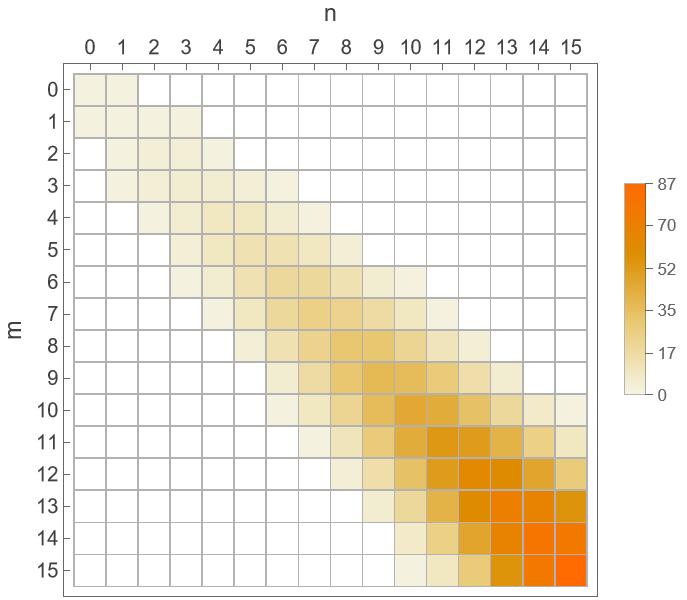} 
\par\end{centering}
\caption{The horizontal axis denotes the powers of $\Lambda_-^{2}$, while the vertical one represents the powers of $\Lambda_+^{2}$. The colored squares correspond to corrections that satisfy \eqref{nonzero}, hence give non-zero contribution to the strong coupling expansion. Different colors represent how many different contributions they contain from the $(\delta^+,\delta^-)$ representation.}
\end{figure}

For the non-zero contributions we also found that at order $\Lambda_-^{2n}\Lambda_+^{2m}$, the leading term in $1/g$ is of order:
\begin{equation}\label{Z-1/g}
    \mathcal{Z}^{(n,m)}(g)=\mathcal{O}(1/g^{(n-m)^2})\,.
\end{equation}
Furthermore, if $n\neq m$, the functions $\mathcal{Z}^{(n,m)}(g)$ are complex valued. Each of these contributions gain an overall prefactor of $e^{i\pi (n-m)a}$. As it can be seen, the $1/g$ powers of the leading order contribution in $\mathcal{Z}^{(n,m)}(g)$ depends on the combination $n-m$. Also the scale $\Lambda_-^{2n}\Lambda_+^{2m}$ gives an additional factor of $1/g^{2a(n-m)}$. Later we will see, that the difference $n-m$ appears in several other ways in describing the large-$g$ expansion of the determinant, therefore it is convenient to introduce the difference:
\begin{equation}\label{Delta-nm}
    \Delta=n-m\,,
\end{equation}
that is different for different non-perturbative sectors. Collecting the leading power of $g$ in each non-perturbative contribution, we find:
\begin{equation}
    \Lambda_-^{2n}\Lambda_+^{2m}\mathcal{Z}^{(n,m)}(g)\sim\mathcal{O}\left(g^{-\Delta(\Delta-2a)}\right)\,.
\end{equation}

As mentioned previously, in addition to the $\alpha\to -\alpha$ symmetry of the determinant, it is also periodic in $\alpha\to \alpha+\pi$. Therefore, it is enough to restrict the values of $\alpha$ to the interval $0<\alpha<\pi/2$. However, for further reasons, it is worth commenting on the results if $\alpha$ is outside of this branch.

The above results are, strictly speaking, only valid in the branch $-\pi/2\leq \alpha \leq \pi/2$. Outside this regime a more careful analysis is necessary. First, consider the modified symbol defined in \eqref{Omega}. This function has the following properties:
\begin{equation}
    \chi_{\alpha}(x)=\chi^*_{-\alpha}(x)=-\chi_{\alpha+\pi}(x)=\chi_{\pi-\alpha}^*(x)\,,
\end{equation}
where "${}^*$" means complex conjugation. The moments \eqref{In} have to reflect these symmetries, namely $\mathcal I_n$ has to satisfy:
\begin{equation}\label{I-prop}
    \mathcal{I}_n(a)=(-1)^n\mathcal{I}_n(-a)=\mathcal{I}_n(a+1)=(-1)^n\mathcal{I}_n(1-a)\,.
\end{equation}
Although the first property is automatically satisfied by the evaluated form \eqref{In-def} of $\mathcal{I}_n$, the function does not admit the rest of the properties. Since \eqref{In-def} has singularities at $a=\pm 1/2$, this problem can be solved by defining $\mathcal{I}_n$ outside the branch $-1/2<a<1/2$ by:
\begin{equation}\label{I-period}
    \mathcal{I}_{n}(a+m)\equiv\mathcal{I}_n(a)\,,
\end{equation}
with some integer $m$. This ensures that the periodicity in \eqref{I-prop} holds.

Since the strong coupling expansion of $Z_\ell(g)$ does not depend on periodic functions of $a$, the remaining explicit dependence on $a$ in the transseries should be shifted according to the branch in which $a$ lies. This, together with the periodicity \eqref{I-period} can be formulated as follows: suppose that for $-1/2<a<1/2$, the strong coupling expansion of $Z_\ell(g)$ is a function of the integrals $\mathcal{I}_n$:
\begin{equation}\label{principal-br}
    Z_\ell(g)\equiv Z_\ell^0[\mathcal I_n](g)\,,
\end{equation}
and it also has an explicit dependence on the parameter $a$. Then for $m-1/2<a<m+1/2$, the strong coupling expansion is given by:
\begin{equation}\label{branch}
    Z_\ell(g)=\left.Z_\ell^0[\mathcal{I}_n]\right|_{a\to a-m}(g)\,.
\end{equation}
That is, the large-$g$ expansion is the same as in the principal branch with the same moments given in \eqref{In-def}, but the {\it explicit} dependence on $a$ is shifted according to the branch of $a$. To distinguish it from the full $a$ dependence (such as in \eqref{Z-symm}), I will use the above notation to indicate that the shift acts only on the explicit $a$ dependence and the moments treated as independent quantities. Later we will see, that the same property appears in the relation connecting the perturbative series with the non-perturbative corrections.

One final remark is that although it is formally possible to determine the transseries \eqref{eq:trans_gamma} up to arbitrary order in $1/g$ and $\Lambda_\pm^2$ with our previous method established in \cite{Bajnok20252}, however, due to the scaling \eqref{Z-1/g}, beyond a certain exponential order, it becomes difficult to gain analytical results for the strong coupling coefficients. Following the idea of \cite{bajnok20253}, in the next section I will show, that \eqref{eq:trans_gamma} can be recast into a new form, in which a natural connection between the perturbative and non-perturbative contributions emerges, making the computation of non-perturbative corrections extremely efficient.

\section{The transseries structure from a different angle}

The reason for the cancelation of terms that do not satisfy the condition in \eqref{nonzero} remained unknown, as it does not follow trivially from the method presented in \cite{Bajnok20252}. It can also be seen that, by applying certain transformations to the parameter $a$ and the integrals $\mathcal I_n$, the non-perturbative functions $\mathcal{Z}^{(n,m)}(g)$ with $(n,m)\neq (0,0)$ can easily be obtained from the perturbative part $\mathcal{Z}^{(0,0)}(g)$. In the following I will show that the procedure of \cite{bajnok20253} can be generalized to the matrix Bessel kernel \eqref{K-alpha} with symbol $\eqref{chi-orig}$, and these properties follow automatically from the analytic structure of the modified symbol \eqref{Omega} and its successive redefinition at each exponential level.

In \cite{Beccaria:2022ypy} it was shown that for $\alpha=0$ and arbitrary symbol function $\chi(x)$, the non-perturbative coefficients for \eqref{F-def} appear at exponential orders $e^{-8\pi g x_n}$, where $x_n$ denote the zeros of the function $1-\chi(x)$. As follows from our method in \cite{Bajnok20252}, in case of the matrix Bessel kernel with mixing angle $\alpha$, this function is replaced by the modified symbol \eqref{Omega}. The solution of the underlying integro-differential equations involves the Wiener-Hopf decomposition of $\chi_\alpha(x)$, which can be written as:
\begin{align}\label{WH}
{}&  \chi_\alpha(x)= {2\cos\pi a \over x} \Phi_+(x)\Phi_{-}(-x) \,,
\end{align}
with $\Phi_\pm(x)$ are given by:
\begin{align}\notag
{}& \Phi_+(x)=\frac{\Gamma \left(\frac12-a\right) \Gamma \left(1+\frac{i x}{2 \pi }\right)}{\Gamma \left(\frac12-a+{ix\over 2\pi}\right)}=\prod_{j\geq0}\frac{1+\frac{ix}{2\pi x_j^+}}{1+\frac{ix}{2\pi y_j}}\,,\\
{}& \Phi_-(x)=\frac{\Gamma \left(\frac12+a\right) \Gamma \left(1+\frac{i x}{2 \pi }\right)}{\Gamma \left(\frac12+a+{ix\over 2\pi}\right)}=\prod_{j\geq0}\frac{1+\frac{ix}{2\pi x_j^-}}{1+\frac{ix}{2\pi y_j}}\,.
\end{align}
These functions are both analytic in the lower half-plane and have infinitely many zeros and poles in the upper half-plane. They vanish at $x=2\pi i x_j^+$ and $x=2\pi i x_j^-$ respectively, with $x_j^\pm$ given by:
\begin{equation}\label{zeros}
    x_j^\pm=j+\frac12\mp a,\qquad j\in \mathbb{N}_0\,,    
\end{equation}
while both of the functions have poles at $x=2\pi i y_j$, where $y_j=j+1$ and $j $ is a non-negative integer.

The functions $\Phi_\pm(x)$ and the zeros $x_j^\pm$ depend on the parameter $a$ and $\Phi_+$ and $x_n^+$ are related to $\Phi_-$ and $x_n^-$ via the exchange $a\to -a$. However, for reasons explained later, I suppress their $a$ dependence and treat these quantities as if they were $a$ independent.

From the method of \cite{Bajnok20252} and the structure studied in \cite{bajnok20253}, we expect the appearance of non-perturbative contributions at orders of $e^{-8\pi g x^\pm_j}$. If we look at the exponential scales $\Lambda_-^{2n}\Lambda_+^{2m}$ multiplying $\mathcal{Z}^{(n,m)}(g)$ in the transseries \eqref{eq:trans_gamma}, we find that the exponential prefactors are:
\begin{equation}\label{expL-nm}
    \Lambda_-^{2n}\Lambda_+^{2m}\sim\exp\left\{-8\pi g \left[n\left(\frac{1}{2}-a\right)+m\left(\frac{1}{2}+a\right)\right]\right\}\,.
\end{equation}
For different values of $n$ and $m$, the exponent contains both $+a$ and $-a$ terms and can be written as finite linear combinations of the zeros $x_j^\pm$, with each $x_j^\pm$ appearing with a multiplicity of one. Therefore, together with the findings of \cite{bajnok20253}, this suggests that in the case of matrix Bessel kernel, non-perturbative contributions contain exponential factors that are at most first order in each parameter $e^{-8\pi g x_j^+}$ and $e^{-8\pi gx_j^-}$. Therefore, the exponential orders can be parametrized as:
\begin{equation}\label{exp-delta}
    e^{-8\pi g \sum_{l\in \delta^+}  x_l^+}e^{-8\pi g \sum_{j\in \delta^- } x_j^-}\,,
\end{equation}
where $\delta^+$ and $\delta^-$ are finite sets of non-negative integers, both containing any integer at most once.

To be more precise, by equating:
\begin{equation}\label{compare}
    \sum_{l\in\delta^+} x_l^++\sum_{j\in \delta^-} x_j^-=n(1/2-a)+m(1/2+a)\,,
\end{equation}
and using \eqref{zeros}, it can be shown that to a specific combination of the sets $\delta^+$ and $\delta^-$, a pair of integers $(n,m)$ can be uniquely assigned, so that \eqref{compare} is satisfied and \eqref{exp-delta} contains the same exponential factor as $\Lambda^{2n}_-\Lambda^{2m}_+$. For example:
\begin{align}\label{delta-nm-eg}
   &\delta^+=\{0\},\,\delta^-=\{\}\ &&\rightarrow& n=1,m=0\,,\notag\\
    &\delta^+=\{0\},\,\delta^-=\{0\}\ &&\rightarrow& n=1,m=1\,,\notag\\
    &\delta^+=\{1\},\,\delta^-=\{\}\ &&\rightarrow& n=2,m=1\,,\notag\\
    &\delta^+=\{0\},\,\delta^-=\{1\}\ &&\rightarrow& n=2,m=2\,,\notag\\
    &\delta^+=\{1\},\,\delta^-=\{0\}\ &&\rightarrow& n=2,m=2\,,\notag\\
    &\delta^+=\{0,1\},\,\delta^-=\{\}\ &&\rightarrow& n=3,m=1\,,\notag\\
    &\delta^+=\{0,1\},\,\delta^-=\{0\}\ &&\rightarrow& n=3,m=2\,,\notag\\
    &\delta^+=\{2\},\,\delta^-=\{\}\ &&\rightarrow& n=3,m=2\,,\notag\\
    &\delta^+=\{1\},\,\delta^-=\{1\}\ &&\rightarrow& n=3,m=3\,.
\end{align}

By investigating further combinations of $\delta^+$ and $\delta^-$ and assigning pairs of integers $(n,m)$ to them via \eqref{compare}, we find that not every pair of $(n,m)$ occurs (eg. $(n,m)\neq(0,2)$). In fact, we only find a contribution at order $\Lambda_-^{2n},\Lambda_+^{2m}$, if $n$ and $m$ satisfy the condition \eqref{nonzero}, so at the same orders as indicated in Figure 1. This confirms that instead of \eqref{eq:trans_gamma}, it is more convenient to parametrize the strong coupling expansion of the determinant in terms of \eqref{exp-delta}.

From the examples in \eqref{delta-nm-eg}, it can also be seen that different combinations of $\delta^+$ and $\delta^-$ could produce the same pair of $(n,m)$. Indeed, by investigating higher order contributions, the number of exponential corrections in \eqref{exp-delta} correspond to the same order $\Lambda_-^{2n}\Lambda_+^{2m}$ rapidly increases. In Figure 1 I indicated the order of this degeneracy using different colors.

It is easy to see that for different sets $(\delta^+,\delta^-)$ that correspond to the same pair of $(n,m)$, the number $\Delta$ defined in \eqref{Delta-nm} is invariant and given by:
\begin{equation}\label{Delta}
    \Delta=n-m=|\delta^+|-|\delta^-|\,.
\end{equation}
Here, $|\delta|$ denotes the number of elements in the set $\delta$. Therefore, the value of $\Delta$ represents the difference between the number of exponential corrections of type $e^{-8\pi g x^+_l}$ and $e^{-8\pi g x^-_j}$.

Finally, collecting all the properties discussed above and in Section \ref{sec:Strong-results}, I conjecture that the strong coupling expansion of $Z_\ell(g)$ is given by:
\begin{align}\label{trans-D}
    Z_\ell(g)=A_\ell(g)&\sum_{\delta^+,\delta^-}(8\pi g)^{-\Delta(\Delta-2a)}e^{-8\pi g \left(\sum_{l\in \delta^+}  x_l^++\sum_{j\in \delta^- } x_j^-\right)}e^{i\pi a\Delta }S^{(\delta^+,\delta^-)}\mathcal{D}^{(\delta^+,\delta^-)}(g)\,.
\end{align}
This series is an alternative form of the strong coupling expansion in \eqref{eq:trans_gamma}. Here, the exponential weights are precisely given by the zeros of the relevant symbol $\chi_\alpha(x)$ and each exponential correction is at most first order in every $e^{-8\pi g x_l^+}$ and $e^{-8\pi g x_j^-}$ factor.

The functions $\mathcal{D}^{(\delta^+,\delta^-)}(g)$ are given by series in $1/g$:
\begin{equation}\label{D-1/g}
    \mathcal{D}^{(\delta^+,\delta^-)}(g)=\sum_{k\geq 0}\frac{d_k^{(\delta^+,\delta^-)}}{(8\pi g)^k}\,,
\end{equation}
and each term in the series \eqref{trans-D} is normalized in such a way that $d_0^{(\delta^+,\delta^-)}=1$ and $S^{(\{\},\{\})}=1$. $S^{(\delta^+,\delta^-)}$ are the Stokes constants of the transseries. Notice that in each term, I have pulled out the complex factors $e^{i\pi a\Delta}$, so that $S^{(\delta^+,\delta^-)}$ and $\mathcal D^{(\delta^+,\delta^-)}(g)$ are real.

The $a\to -a$ symmetry of the determinant is reflected in the Stokes constants and the non-perturbative functions as:
\begin{align}\notag\label{DS-symm}
    S^{(\delta^+,\delta^-)}(a)&=S^{(\delta^-,\delta^+)}(-a)\,,\notag\\
    \mathcal D^{(\delta^+,\delta^-)}(g|a)&=\mathcal D^{(\delta^-,\delta^+)}(g|-a)\,.
\end{align}

In the perturbative sector $\delta^{\pm}=\{\}$ the coefficients $d_k^{(\{\},\{\})}$ with $k\geq 1$ are related to the functions $f_k^{(0,0)}$ given in \eqref{f00} via:
\begin{equation}\label{d-f-rel}
    d_k^{(\{\},\{\})}=(a^2-\ell^2)\frac{(-1)^k}{k!}f_k^{(0,0)}\,.
\end{equation}

For the leading order non-perturbative contribution $\delta^+=\{0\}$ and $\delta^-=\{\}$ we have:
\begin{equation}
    d_k^{(\{0\},\{\})}=((a-1)^2-\ell^2)\frac{(-1)^k}{k!}f_k^{(1,0)}\,,
\end{equation}
with the first few $f_k^{(1,0)}$ are given in \eqref{f10}. At this exponential level the Stokes constant coincides with $P^{(1,0)}$ and by \eqref{P10} its explicity value is:
\begin{equation}
    S^{(\{0\},\{\})}=-(-1)^{\ell }\left(\frac{1}{2}-a\right)^{2 a-1} \frac{ \Gamma (1+\ell-a)}{ \Gamma (\ell+a )}\,.
\end{equation}
The next subleading coefficients $d_k^{(\{\},\{0\})}$ and the corresponding Stokes constant $S^{(\{\},\{0\})}$ can be obtained from the ${(\{0\},\{\})}$ sector by changing $a\to -a$.

For subleading corrections, the situation becomes more complicated. As can be seen in \eqref{delta-nm-eg}, for example, the $\Lambda_-^4\Lambda_+^4$ term receives a contribution from two different sources in the sum \eqref{trans-D}: from the corrections with $\delta^+=\{0\}$, $\delta^-=\{1\}$ and $\delta^+=\{1\}$, $\delta^-=\{0\}$. Therefore, at this order $\mathcal{Z}^{(2,2)}(g)$ is expected to be written as a sum:
\begin{equation}\label{ZDrel}
    \mathcal{Z}^{(2,2)}(g)=S^{(\{1\},\{0\})}\mathcal{D}^{(\{1\},\{0\})}(g)+S^{(\{0\},\{1\})}\mathcal{D}^{(\{0\},\{1\})}(g)\,.
\end{equation}
 
In the following I present a method to compute the non-perturbative functions $\mathcal{D}^{(\delta^+,\delta^-)}(g)$ from the perturbative one $\mathcal{D}^{(\{\},\{\})}$. By comparing numerically the results with the series \eqref{eq:trans_gamma} obtained in \cite{Bajnok20252}, I verify the degeneracy of higher order contributions $\mathcal{Z}^{(n,m)}(g)$ in the $\delta^\pm$ prescription. I also show a recursive way to effectively generate the Stokes constants $\mathcal{S}^{(\delta^+,\delta^-)}$. This fully describes the strong coupling expansion of the determinant \eqref{F-def}.

\subsection{$1/g$ expansion}

First, I describe the structure of the series $\mathcal{D}^{(\delta^+,\delta^-)}(g)$. As was discussed in Section \ref{sec:Strong-results}, the $1/g$ expansion of these functions is governed by the integrals $\mathcal{I}_n$. Their values are given in \eqref{In-def}. Using the series representation of the polygamma function, these integrals can be directly expressed in terms of the zeros and poles of the the symbol $\chi_\alpha(x)$ (or $\Phi_\pm(x)$):
\begin{equation}\label{I-zp}
    \mathcal I_n=\sum_{k=0}^\infty\left[(-1)^{n-1}\left(\frac{1}{(x_k^+)^{n-1}}-\frac{1}{(y_k)^{n-1}}\right)-\left(\frac{1}{(x_k^-)^{n-1}}-\frac{1}{(y_k)^{n-1}}\right)\right]\,.
\end{equation}
the locations of the zeros $x_k^\pm$ are given in \eqref{zeros}. 

In Section \ref{sec:Strong-results}, I have shown, that while the perturbative series $\mathcal{D}^{(\{\},\{\})}$ is given in terms of \eqref{I-zp}, the $1/g$ expansions of the leading order non-perturbative sectors $\mathcal{Z}^{(1,0)}(g)$, $\mathcal{Z}^{(0,1)}(g)$ and $\mathcal{Z}^{(1,1)}(g)$ -- therefore $\mathcal{D}^{(\{0\},\{\})}(g)$, $\mathcal{D}^{(\{\},\{0\})}(g)$ and $\mathcal{D}^{(\{0\},\{0\})}(g)$ -- can be easily obtained from the perturbative series by applying the transformations \eqref{shift-gen-10} and \eqref{shift-gen-01} and shifting the explicit dependence on $a$ by $a\to a-1$ and $a\to a+1$ and $a\to a$ respectively. In terms of the zeros of $\Phi_\pm(x)$, for example, the transformation \eqref{shift-gen-10} is simply written as:
\begin{equation}
    \mathcal{I}_n\to \mathcal{I}^{(\{0\},\{\})}_n\equiv \mathcal{I}_n-(-1)^{n-1}\frac{2}{\left(x_0^+\right)^{n-1}}\,.
\end{equation}
From the point of view of expansion \eqref{I-zp}, this rule corresponds to changing the sign before the terms containing $x_0^+$ to contribute to $\mathcal{I}_n$ in the same way as the locations of the poles $y_k$ do. In other words, the "new" integral $\mathcal{I}^{(\{0\},\{\})}_n$ entering the non-perturbative series $\mathcal{Z}^{(1,0)}(g)$ is given by the same integral as in \eqref{In}, but with a new symbol, $\chi^{(\{0\},\{\})}_\alpha(x)$:
\begin{equation}
    \chi^{(\{0\},\{\})}_\alpha(x)={2\cos\pi a \over x} \Phi^{\{0\}}_+(x)\Phi_{-}(-x)\,,
\end{equation}
where the function $\Phi^{\{0\}}_+(x)$ is:
\begin{equation}
    \Phi^{\{0\}}_+(x)=\frac{\Phi_+(x)}{\left(1+\frac{ix}{2\pi x_0^+}\right)^2}\,.
\end{equation}
This function has zeros at $x=2i\pi x_j^+$ with $j\geq 1$ and poles at $x=2i\pi x^+_0$ and $x=2i\pi y_j$ with $j\geq 0$.

We can interpret the transformation \eqref{shift-gen-01} in the same way, but with another symbol:
\begin{equation}
    \chi^{(\{\},\{0\})}_\alpha(x)={2\cos\pi a \over x} \Phi_+(x)\Phi^{\{0\}}_{-}(-x)\,,
\end{equation}
with the function $\Phi^{\{0\}}_{-}(x)$ being:
\begin{equation}
    \Phi^{\{0\}}_-(x)=\frac{\Phi_-(x)}{\left(1+\frac{ix}{2\pi x_0^-}\right)^2}\,.
\end{equation}
Now, this function has zeros at $x=2i\pi x_j^-$ with $j\geq 1$ and poles at $x=2i\pi x^-_0$ and $x=2i\pi y_j$ with $j\geq 0$. These transformation rules are analogous to what we have found in \cite{bajnok20253}.

This suggests that all the subleading non-perturbative corrections can be obtained in a similar way, by promoting the corresponding zeros to be poles of the functions $\chi_\alpha(x)$, namely at every exponential level introducing a new symbol function $\chi^{(\delta^+,\delta^-)}$:
\begin{equation}
    \chi_\alpha^{(\delta^+,\delta^-)}(x)={2\cos\pi a \over x} \Phi^{\delta^+}_+(x)\Phi^{\delta^-}_{-}(-x)\,,
\end{equation}
with:
\begin{align}\label{phi-new}
    \Phi_+^{\delta^+}(x)&=\frac{\Phi_+(x)}{\prod_{j\in\delta^+}\left(1+\frac{ix}{2\pi x_j^+}\right)^{2}}\,,&&
    \Phi_-^{\delta^-}(x)=\frac{\Phi_-(x)}{\prod_{j\in \delta^-}\left(1+\frac{ix}{2\pi x_j^-}\right)^{2}}\,.
\end{align}
This new symbol enters the $1/g$ expansion of $\mathcal{D}^{(\delta^+,\delta^-)}$ through the integrals of the form \eqref{In} but with $\chi_\alpha(x)$ replaced by $\chi^{(\delta^+,\delta^-)}_\alpha(x)$. This results in the modified moments:
\begin{align}\label{I-mod-exact}
    \mathcal{I}_n^{(\delta^+,\delta^-)}&=\mathcal{I}_n-(-1)^{n-1}\sum_{l\in\delta^+}\frac{2}{(x_l^+)^{n-1}}+\sum_{j\in\delta^-}\frac{2}{(x_j^-)^{n-1}}=\notag\\
    &=\mathcal{I}_n-(-1)^{n-1}\sum_{l\in\delta^+}\frac{2}{\left(l+\frac{1}{2}-a\right)^{n-1}}+\sum_{j\in\delta^-}\frac{2}{(j+\frac{1}{2}+a)^{n-1}}\,,
\end{align}
where in the second relation, I restated the exact form of $x^\pm_l$. Taking into account all the properties discussed above, the function $\mathcal{D}^{(\delta^+,\delta^-)}(g)$ is given by the perturbative function by replacing the integrals $\mathcal I_n$ by \eqref{I-mod-exact} and shifting the explicit $a$ dependence by $a\to a-\Delta$, with $\Delta$ defined in \eqref{Delta}

In other words, if we indicate the functional dependence of the perturbative part on $\mathcal{I}_n$:
\begin{equation}
    \mathcal{D}^{(\{\},\{\})}(g)\equiv \mathcal{D}\left[\mathcal{I}_n\right]( g)\,,
\end{equation}
then the non-perturbative contribution $\mathcal{D}^{(\delta^+,\delta^-)}(\tilde g)$ is given by:
\begin{equation}\label{D-rule}
    \mathcal{D}^{(\delta^+,\delta^-)}( g)=\left.\mathcal{D}\left[\mathcal{I}^{(\delta^+,\delta^-)}_n\right]( g)\right|_{a\to a-\Delta}\,.
\end{equation}

To put this in an explicit form, using \eqref{Z00} an \eqref{d-f-rel}, we can parameterize any non-perturbative function $\mathcal{D}^{(\delta^+,\delta^-)}(g)$ as:
\begin{equation}
    \mathcal{D}^{(\delta^+,\delta^-)}(g)=1+\left((a-\Delta)^2-\ell^2\right)\sum_{k\geq 1}\frac{(-1)^k}{k!}\frac{f_k^{(\delta^+,\delta^-)}}{(8\pi g)^k}\,,
\end{equation}
with the coefficients $f^{(\delta^+,\delta^-)}_k$ given by:
\begin{align}\notag\label{f-nonP}
    f^{(\delta^+,\delta^-)}_1&=\mathcal{I}^{(\delta^+,\delta^-)}_2\,,\\
    f^{(\delta^+,\delta^-)}_2&=((a-\Delta)^2-\ell^2+1)\left[\mathcal{I}^{(\delta^+,\delta^-)}_2\right]^2+2(a-\Delta)\mathcal{I}^{(\delta^+,\delta^-)}_3\,,\notag\\
    f^{(\delta^+,\delta^-)}_3&=\left((a-\Delta)^2-\ell ^2+1\right) \left[\mathcal{I}^{(\delta^+,\delta^-)}_2\right]^3+\notag\\
    &+6 (a-\Delta) \left((a-\Delta)^2-\ell ^2+2\right)\mathcal{I}^{(\delta^+,\delta^-)}_2\mathcal{I}^{(\delta^+,\delta^-)}_3+\notag\\
    &+2(5 (a-\Delta)^2- \ell ^2+1)\mathcal{I}^{(\delta^+,\delta^-)}_4\,,
\end{align}
etc.

With this simple rule, up to the overall factor $S^{(\delta^+,\delta^-)}$, it is possible to generate any non-perturbative contributions to the determinant \eqref{F-def}, by computing the perturbative series of a "new" determinant with a suitably modified symbol function. 

At the end of Section 2, I discussed the effect of the periodicity of $Z_\ell(g)$ in $\alpha\to \alpha+\pi$ on the strong coupling expansion. According to \eqref{principal-br} and \eqref{branch}, the right-hand side of \eqref{D-rule} is the large-$g$ perturbative series of the determinant evaluated in the branch $\Delta-1/2<a<\Delta+1/2$ with the original symbol $\chi_\alpha(x)$ replaced by $\chi^{(\delta^+,\delta^ -)}_\alpha(x)$.

It is important to emphasize that although the zeros $x_j^\pm$ and hence the integrals $\mathcal{I}_n$ are $a$ dependent, the shift $a\to a-\Delta$ in \eqref{D-rule} only stands for the explicit $a$ dependence in the coefficients $f_k^{(0,0)}$. This is the reason why I have suppressed the $a$ dependence in the notation $x^\pm_j$ and $\mathcal{I}_n$.

For the parameter values of $(a,\ell)=(1/4,0)$, $(a,\ell)=(1/4,1)$, and $(a,\ell)=(1/(2\sqrt{2}),2)$, I numerically generated 50 $1/g$ terms in the expansions of $\mathcal{D}^{(\{0\},\{\})}(g)$, $\mathcal{D}^{(\{\},\{0\})}(g)$ and $\mathcal{D}^{(\{0\},\{0\})}(g)$ with 50 digit precision. By normalizing the functions $\mathcal{Z}^{(n,m)}(g)$ with their leading $1/g$ coefficients, I was able to verify \eqref{D-rule} for these non-degenerate contributions. For the next few degenerate non-perturbative sectors, from the analytic values of their leading $1/g$ coefficients, I checked that there exist certain overall constant factors $S^{(\delta^+,\delta^-)}$, such that relations like \eqref{ZDrel} hold, with $\mathcal{D}^{(\delta^+,\delta^-)}(g)$ determined via \eqref{D-rule}.

\subsection{Stokes constants}

The only remaining task to completely describe the large-$g$ expansion of the determinant is to give a closed formula for the Stokes constants $S^{(\delta^+,\delta^-)}$. In the following I show that there is a consistent prescription to determine these constants with the argument given for $\mathcal{D}^{(\delta^+,\delta^-)}( g)$ above, which allows to derive a pair of recurrence relations for the Stokes constants.

I start by rewriting the leading order constants $S^{(\{0\},\{\})}$, $S^{(\{\},\{0\})}$ and $S^{(\{0\},\{0\})}$ -- which, according to \eqref{delta-nm-eg}, are essentially the same as $P^{(1,0)}$, $P^{(0,1)}$ and $P^{(1,1)}$ --   in terms of the functions $\Phi_\pm(x)$ and the locations of their zeros and poles $x_j^\pm$ and $y_j$. From the method presented in \cite{Bajnok20252} it follows that the factors $S^{(\{0\},\{\})}$ and $S^{(\{\},\{0\})}$ can be written as:
\begin{align}\notag\label{S-leading}
    S^{(\{0\},\{\})}&=\frac{(-1)^\ell}{2\pi i}\,\frac{ \Gamma(1+\ell-a)}{\Gamma(\ell+a)}\frac{\Phi_{-}(-2i\pi x_0^+)}{\partial_x\Phi_{+}(2i \pi x_0^+)}(x_0^+)^{2a-2}\,,\notag\\
    S^{(\{\},\{0\})}&=\frac{(-1)^\ell}{2\pi i}\,\frac{\Gamma(1+\ell+a)}{\Gamma(\ell-a)}\frac{\Phi_{+}(-2i\pi x_0^-)}{\partial_x\Phi_{-}(2i\pi x_0^-)}(x_0^-)^{-2a-2}\,.
\end{align}
By investigating higher order contributions, it turns out that similar combinations appear in the Stokes constants at each exponential level, but with $x_0^\pm$ replaced by $x_j^\pm$. Therefore, it is convenient to introduce the functions:
\begin{align}\label{Selem}
    S_+^{\{j\}}&=\frac{(-1)^\ell}{2\pi i}\,\frac{ \Gamma(\ell-a)}{\Gamma(1+\ell+a)}\frac{\Phi_{-}(-2i\pi x_j^+)}{\partial_x\Phi_{+}(2i \pi x_j^+)}(x_j^+)^{2a}\,,\notag\\
    S_-^{\{j\}}&=\frac{(-1)^\ell}{2\pi i}\,\frac{\Gamma(\ell+a)}{\Gamma(1+\ell-a)}\frac{\Phi_{+}(-2i\pi x_j^-)}{\partial_x\Phi_{-}(2i\pi x_j^-)}(x_j^-)^{-2a}\,.
\end{align}
Then the leading order coefficients are given by:
\begin{align}\notag\label{S-10}
    S^{(\{0\},\{\})}&=\left.S^{\{0\}}_+\right|_{a\to a-1}\,,\\
 S^{(\{\},\{0\})}&=\left.S^{\{0\}}_-\right|_{a\to a+1}\,.
\end{align}
Notice that the shift in $a$ is again understood only for the explicit dependence on $a$ and the functions $\Phi_\pm(x)$ and their zeros $x^\pm_j$ are treated as independent objects. $S_\pm^{\{j\}}$ was chosen in such a way that this shift is consistent with $a\to a-\Delta$ appearing in \eqref{D-rule}. 

By generating exact analytical expressions for subleading corrections up to $\Lambda_-^{6}\Lambda_+^{6}$, I found that at orders $\mathcal{O}\left(e^{-8\pi g x_j^\pm}\right)$, similar relations to \eqref{S-10} still hold, namely at orders $(\delta^+,\delta^-)=(\{j\},\{\})$, the Stokes constants are:
\begin{align}\label{Spj}
    S^{(\{j\},\{\})}=\left.S^{\{j\}}_+\right|_{a\to a-1}\,,
\end{align}
whereas if $\delta^+=\{\}$ and $\delta^-=\{j\}$ we have:
\begin{align}\label{Smj}
    S^{(\{\},\{j\})}=\left.S^{\{j\}}_-\right|_{a\to a+1}\,.
\end{align}

The problem arises for mixed terms, that is, when $\delta^+$ and $\delta^-$ together contain more than one element. By generating subleading corrections $\mathcal{D}^{(\delta^+,\delta^-)}(g)$ with \eqref{D-rule} and, with their proper degeneracy, comparing their linear combinations with $\mathcal{Z}^{(\delta^+,\delta^-)}$, for the first few subleading Stokes constants, I found:
\begin{align}\notag\label{Stokes-exam}
    &S^{(\{0\},\{0\})}=-\left(\frac{1}{2}-a\right)^{1+2a}\left(\frac{1}{2}+a\right)^{1-2a}\,,\notag\\
    &S^{(\{0,1\},\{\})}=-\left(\frac{1}{2}-a\right)^{-1+2a}\left(\frac{3}{2}-a\right)^{-3+2a}\frac{\Gamma (\ell-a+1) \Gamma (\ell-a+2)}{\Gamma (\ell +a)\Gamma (\ell+a -1) }\,,\notag\\
    &S^{(\{0,1\},\{0\})}=\frac{(-1)^\ell}{4}\left(\frac{1}{2}+a\right)^{3-2a}\left(\frac{1}{2}-a\right)^{1+2a}\left(\frac{3}{2}-a\right)^{-1+2a}\frac{\Gamma(\ell-a+1)}{\Gamma(\ell+a)}\,,\notag\\
    &S^{(\{0,1,2\},\{\})}=\left(\frac{1}{2}-a\right)^{-1+2a}\left(\frac{3}{2}-a\right)^{-3+2a}\left(\frac{5}{2}-a\right)^{-5+2a}\frac{\Gamma (\ell-a+1) \Gamma (\ell-a+2)\Gamma(\ell-a+3)}{ \Gamma (\ell +a)\Gamma (\ell+a -1)\Gamma(\ell+a-2)}\,,\notag\\
    &S^{(\{0,1\},\{0,1\})}=\frac{1}{144}\left(\frac{1}{2}-a\right)^{3+2a}\left(\frac{3}{2}-a\right)^{1+2a}\left(\frac{1}{2}+a\right)^{3-2a}\left(\frac{3}{2}+a\right)^{1-2a}\,.
\end{align}
Further can be obtained by using the symmetry property in \eqref{DS-symm}.

The first two coefficients above (together with their $a\to -a$ counterparts) can also be expressed in terms of the functions in \eqref{Selem}:
\begin{align}\notag\label{SsubV1}
    S^{(\{0,1\},\{\})}&=-\left(1-\frac{x_1^+}{x_0^+}\right)^2\left(\left.S_+^{\{0\}}\right|_{a\to a-1}\right)\left(\left.S_+^{\{1\}}\right|_{a\to a-2}\right)\,,\notag\\
    S^{(\{\},\{0,1\})}&=-\left(1-\frac{x_1^-}{x_0^-}\right)^2\left(\left.S_-^{\{0\}}\right|_{a\to a+1}\right)\left(\left.S_-^{\{1\}}\right|_{a\to a+2}\right)\,,\notag\\
    S^{(\{0\},\{0\})}&=-\left(1+\frac{x_0^-}{x_0^+}\right)^{-2}\left(\left.S_+^{\{0\}}\right|_{a\to a-1}\right)S_-^{\{0\}}\,.  
\end{align}
Therefore, in addition to the products of $S^{\{j\}}_\pm$ with certain shifts in parameter $a$, we have additional factors containing the zeros $x_j^\pm$. Moreover, for the mixed terms, there is more than one way to express them in terms of \eqref{Selem}. For example, beside \eqref{SsubV1}, we can also write:
\begin{align}\notag\label{SsubV2}
    S^{(\{0,1\},\{\})}&=-\left(1-\frac{x_0^+}{x_1^+}\right)^2\left(\left.S_+^{\{1\}}\right|_{a\to a-1}\right)\left(\left.S_+^{\{0\}}\right|_{a\to a-2}\right)\notag\,,\\
    S^{(\{\},\{0,1\})}&=-\left(1-\frac{x_0^-}{x_1^-}\right)^2\left(\left.S_-^{\{1\}}\right|_{a\to a+1}\right)\left(\left.S_-^{\{0\}}\right|_{a\to a+2}\right)\notag\,,\\
    S^{(\{0\},\{0\})}&=-\left(1+\frac{x_0^+}{x_0^-}\right)^{-2}\left(\left.S_-^{\{0\}}\right|_{a\to a+1}\right)S_+^{\{0\}}\,. 
\end{align}
In this case, there are different shifts in $a$ and different prefactors, but surprisingly the results are the same as \eqref{SsubV2}.

Following the same argument that I presented for the functions $\mathcal D^{(\delta^+,\delta^-)}( g)$, these properties can be understood in a simple way. Assume that we start with the exponential correction $(\delta^+,\delta^-)=(\{j\},\{\})$ whose Stokes constant is given in \eqref{Spj}. Now append an element $k\neq j$ to $\delta^+$, so we have the exponential correction at level $(\delta^{'+},\delta^-)=(\{j,k\},\{\})$. With this step, $\Delta$ increased and became $\Delta'=|\delta^{'+}|-|\delta^-|=2$. Then motivated by the structure of $\mathcal D^{(\delta^+,\delta^-)}(g)$, it is reasonable to guess that the new Stokes constant is \eqref{Spj} multiplied by $S_+^{\{k\}}$ with its explicit $a$ dependence shifted by $a\to a-\Delta'=a-2$. However, in addition to the shift, we saw that at every exponential level, to generate the functions $\mathcal{D}^{(\delta^+,\delta^-)}( g)$, we have to remove the corresponding zeros of $\Phi_\pm(x)$ and promote them to be poles. Therefore, we have to use the new functions in \eqref{phi-new} to describe the $1/g$ expansion. Since at the level $(\delta^+,\delta^-)=(\{j\},\{\})$ we already "removed" the $x_j^+$ zero, this suggests that to obtain $S^{(\{j,k\},\{\})}$ from $S^{(\{j\},\{\})}$, we have to multiply \eqref{Spj} with:
\begin{equation}\label{example-1}
    \left.\frac{(-1)^\ell}{2\pi i}\,\frac{ \Gamma(\ell-a)}{\Gamma(1+\ell+a)}\frac{\Phi_{-}(-2i\pi x_k^+)}{\partial_x\Phi^{\{j\}}_{+}(2i \pi x_k^+)}(x_k^+)^{2a}\right|_{a\to a-2}=\left(1-\frac{x_k^+}{x_j^+}\right)^2\left(\left.S^{\{k\}}_+\right|_{a\to a-2}\right)\,.
\end{equation}
The differentiation in the denominator on the left-hand side acts on the new function $\Phi_+^{\{j\}}(x)$ defined in \eqref{phi-new}. Since $\Phi_+(x)$ in the numerator of \eqref{phi-new} vanishes at $x=2i\pi x_k^+$ for every $k\geq 0$, therefore, the derivative in the denominator simplifies as:
\begin{equation}
    \partial_x\Phi^{\{j\}}_{+}(2i \pi x_k^+)={\partial_x\Phi_{+}(2i \pi x_k^+)\over \left(1-{x_k^+\over x_j^+}\right)^2 }\,,
\end{equation}
which results in the additional overall factor $\left(1-x_k^+/x_j^+\right)^2$ on the right-hand side of \eqref{example-1}. In case of $j=0$, $k=1$ and $j=1$, $k=0$, up to an overall minus sign, this reproduces the first lines of \eqref{SsubV1} and \eqref{SsubV2}.

Similarly if we start with $(\delta^+,\delta^-)=(\{j\},\{\})$ and go to level $(\delta^+,\delta^{'-})=(\{j\},\{k\})$ then we get the same number of "$+$" type contributions as "$-$" ones, therefore at this level we have $\Delta'=|\delta^{+}|-|\delta^{'-}|=0$. In the first step $x_j^+$ was removed from the symbol. Therefore, it suggests that the new Stokes constant is given by the product of \eqref{Smj} and:
\begin{align}
    \frac{(-1)^\ell}{2\pi i}\,\frac{\Gamma(\ell+a)}{\Gamma(1+\ell-a)}\frac{\Phi^{\{j\}}_{+}(-2i\pi x_k^-)}{\partial_x\Phi_{-}(2i\pi x_k^-)}(x_k^-)^{-2a}=\left(1+\frac{x_k^-}{x_j^+}\right)^{-2}S_-^{\{k\}}\,.
\end{align}
With $j=0$ and $k=0$, up to an overall minus sign, this reproduces the result in the last line of \eqref{SsubV1}.

The similar procedure can be done by starting from $(\delta^+,\delta^-)=(\{\},\{j\})$ to reproduce the rest of \eqref{SsubV1} and \eqref{SsubV2}. The simultaneous shifts of $a$ and removal of zeros guarantees that the two prescriptions give the same answer independently of the path of adding elements to $\delta^\pm$.

This procedure can be generalized to arbitrary exponentially small corrections. By appending more and more elements to $\delta^+$ and $\delta^-$, at each step we have to multiply the previous Stokes constant with \eqref{Selem}, but with $\Phi_\pm(x)$ replaced by $\Phi_\pm^{\delta^\pm}(x)$ and the explicit $a$ dependence shifted by the new $\Delta$. 

The procedure can be summarized in the following way: the Stokes constant for the perturbative part is:
\begin{equation}
    S^{(\{\},\{\})}=1\,.
\end{equation}
Now assume that we know the value of $S^{(\delta^+,\delta^-)}$ at a specific level $(\delta^+,\delta^-)$. Then the Stokes constant at level $(\delta^{'+},\delta^-)=(\delta^+\cup\{k\},\delta^-)$, which is obtained from $(\delta^{+},\delta^-)$ by adding the integer $k\notin \delta^+$ to the set $\delta^+$ is given by:
\begin{equation}\label{SrecP}
S^{(\delta^{'+},\delta^{-})}=(-1)^{\Delta}\frac{\prod_{j\in \delta^+}\left(1-\frac{x_k^+}{x_j^+}\right)^{2}}{\prod_{j\in \delta^-}\left(1+\frac{x_k^+}{x_{j}^-}\right)^{2}}\left(\left.S_+^{\{k\}}\right|_{a\to a-\Delta'}\right)S^{(\delta^{+},\delta^{-})}\,,
\end{equation}
where $\Delta=|\delta^+|-|\delta ^-|$ and $\Delta'=|\delta^{'+}|-|\delta ^-|=\Delta+1$.

Similarly, the Stokes constant at level $(\delta^+,\delta^{'-})=(\delta^+,\delta^-\cup\{k\})$, which differs from the original $(\delta^+,\delta^-)$ by adding the element $k\notin \delta^-$ to $\delta^{-}$, is given by:
\begin{align}\label{SrecM}
S^{(\delta^{+},\delta^{'-})}=(-1)^{\Delta}\frac{\prod_{j\in \delta^-}\left(1-\frac{x_k^-}{x_j^-}\right)^{2}}{\prod_{j\in\delta^+}\left(1+\frac{x_k^-}{x_{j}^+}\right)^{2}}\left(\left.S_-^{\{k\}}\right|_{a\to a-\Delta'}\right)S^{(\delta^{+},\delta^{-})}\,,
\end{align}
where in this case $\Delta'=|\delta^{+}|-|\delta ^{'-}|=\Delta-1$. Using these recurrence relations, the Stokes constants can be determined up to arbitrary non-perturbative orders.

The prefactors $(-1)^\Delta$ in \eqref{SrecM} and \eqref{SrecP} were introduced by investigating numerical results with different values of $a$ and $\ell$ up to $\Lambda_-^5\Lambda_+^5$. The origin of these factors can be understood by absorbing the $e^{i\pi a\Delta}$ factors from \eqref{trans-D} as $e^{\pm i\pi a}$ phases in \eqref{Selem}. Then the Stokes constants will no longer be real, but the $a\to a-\Delta$ shifts in the recurrence relations \eqref{SrecP} and \eqref{SrecM} will produce exactly the same factors $(-1)^{\Delta}$ as in \eqref{SrecP} and \eqref{SrecM} and will produce the same complex phases $e^{i\pi a\Delta}$ as in \eqref{trans-D}. We will see later that the prescription that I used above is more convenient regarding the resurgence analysis of the transseries \eqref{trans-D}.

To put the recurrence relations \eqref{SrecP} and \eqref{SrecM} into a more practical form, I recover the explicit forms of the functions $\Phi_\pm(x)$ and the zeros $x^\pm_l$. With this, the first recurrence relation becomes
\begin{align}\notag\label{Sp-fin}
S^{(\delta^{+}\cup\{k\},\delta^-)}=&-(-1)^{\ell+\Delta}\frac{\Gamma\left(\ell-a+\Delta+1\right)}{\Gamma\left(\ell+a-\Delta\right)}\frac{\Gamma^2\left(k+\frac{1}{2}-a\right)}{\Gamma^2\left(\frac{1}{2}-a\right)\Gamma^2\left(k+1\right)}\times\notag\\
&\times\frac{\prod_{j\in \delta^+}\left(\frac{j-k}{j+\frac{1}{2}-a}\right)^{2}}{\prod_{j\in\delta^-}\left(\frac{j+k+1}{j+\frac{1}{2}+a}\right)^{2}}\left(k+\frac{1}{2}-a\right)^{2a-1-2\Delta}S^{(\delta^{+},\delta^{-})}\,,
\end{align}
while the second one is:
\begin{align}\notag\label{Sm-fin}
S^{(\delta^{+},\delta^{-}\cup\{k\})}=&-(-1)^{\ell-\Delta}\frac{\Gamma\left(\ell+a-\Delta+1\right)}{\Gamma\left(\ell-a+\Delta\right)}\frac{\Gamma^2\left(k+\frac{1}{2}+a\right)}{\Gamma^2\left(\frac{1}{2}+a\right)\Gamma^2\left(k+1\right)}\times\notag\\
&\times\frac{\prod_{j\in \delta_j^-}\left(\frac{j-k}{j+\frac{1}{2}+a}\right)^{2}}{\prod_{j\in \delta_j^+}\left(\frac{j+k+1}{j+\frac{1}{2}-a}\right)^{2}}\left(k+\frac{1}{2}+a\right)^{-2a-1+2\Delta}S^{(\delta^{+},\delta^{-})}\,.
\end{align}
These relations reproduce the examples given in \eqref{Stokes-exam}.

The recurrence relations \eqref{SrecP} and \eqref{SrecM} together with \eqref{D-rule} completely describe the large-$g$ expansion of the determinant $\mathcal{Z}(g)$ and allow us to effectively determine the physical observables discussed in Section 1 up to arbitrary non-perturbative and $1/g$ orders by only knowing the perturbative contributions of the determinant. The relations \eqref{D-rule}, \eqref{Sp-fin} and \eqref{Sm-fin} were numerically verified for the parameter values of $(a,\ell)=(1/4,0)$, $(a,\ell)=(1/4,1)$ and $(a,\ell)=(1/(2\sqrt{2}),2)$ by comparing different corrections in \eqref{trans-D} with the corresponding terms \eqref{eq:trans_gamma} up to order $\Lambda_-^8\Lambda_+^8$.

\subsection{A practical example}

In this subsection, I specify the parameter $a$ as $a=1/4$ and generate exact strong-coupling results for the determinant. As was discussed in the Introduction, this is a relevant example for the $\mathcal{N}=4$ SYM, since the ratio of the determinants evaluated at $\ell=1$ and $\ell=0$ gives the cusp anomalous dimension (see equation \eqref{tilted-cusp}).

For $a=1/4$, the zeros of $\Phi_\pm(x)$ are located at:
\begin{equation}
    x^+_j=j+\frac{1}{4}\,,\qquad x_j^-=j+\frac{3}{4}\,,
\end{equation}
with $j\geq0$ an integer. Therefore, in this case, the exponential corrections scale as powers of $e^{-2\pi g}$ and we can parametrize the strong coupling expansion for $a=1/4$ as:
\begin{equation}\label{a=1/4-exp}
    Z_\ell(g)=A_\ell(g)\sum_{n\geq0}e^{-2\pi n g}\mathcal C_\ell^{(n)}(g)\,.
\end{equation}
Using \eqref{D-rule}, the first few coefficient functions $\mathcal C_\ell^{(n)}(g)$ are:
\begin{align}\notag\label{Cs}
    &\mathcal C_\ell^{(0)}(g)=\left.\mathcal{D}^{(\{\},\{\})}(g)\right|_{a=1/4}\,,&&
    \mathcal C_\ell^{(1)}(g)=e^{\frac{i\pi}{4}}(8\pi g)^{-\frac{1}{2}}\left.S^{(\{0\},\{\})}\mathcal{D}^{(\{0\},\{\})}(g)\right|_{a=1/4}\,,\notag\\
    &\mathcal C_\ell^{(2)}(g)=0\,,&&
    \mathcal C_\ell^{(3)}(g)=e^{-\frac{i\pi}{4}}(8\pi g)^{-\frac{3}{2}}\left.S^{(\{\},\{0\})}\mathcal{D}^{(\{\},\{0\})}(g)\right|_{a=1/4}\,,\notag\\
    &\mathcal C_\ell^{(4)}(g)=\left.S^{(\{0\},\{0\})}\mathcal{D}^{(\{0\},\{0\})}(g)\right|_{a=1/4}\,,&&\mathcal C_\ell^{(5)}(g)=e^{\frac{i\pi}{4}}(8\pi g)^{-\frac{1}{2}}\left.S^{(\{1\},\{\})}\mathcal{D}^{(\{1\},\{\})}(g)\right|_{a=1/4}\,.
\end{align}

In the examples above, for a given $\mathcal C_\ell^{(n)}(g)$, only one non-perturbative function $\mathcal D^{(\delta^+,\delta^-)}(g)$ contributes. However, for $a=1/4$, at levels with $n\geq 8$, certain combinations of $(\delta^+,\delta^-)$ produce an exponential contribution to \eqref{a=1/4-exp} at the same level, therefore, in general $\mathcal C^{(n)}(g)$ is a linear combination of multiple $\mathcal{D}^{(\delta^+,\delta^-)}(g)$. At higher orders, to compute $\mathcal{C}^{(n)}(g)$, one should sum up all the $(\delta^+,\delta^-)$ corrections in \eqref{trans-D}, for which:
\begin{equation}\label{degen-cusp}
    \sum_{l\in \delta^+}\left(4l+1\right)+\sum_{j\in \delta^-}\left(4j+3\right)=n\,.
\end{equation}

Since $\mathcal I_n$ are independent of the parameter $\ell$, before giving explicit expressions for the different coefficients in \eqref{a=1/4-exp} with $\ell=0,1$, it is convenient to rewrite the moments in a more familiar form. By substituting $x^\pm_k$ with $a=1/4$ into \eqref{I-zp}, the moments $\mathcal{I}_n$ with odd $n$ can be written as:
\begin{equation}\label{I-odd}
    \mathcal{I}_{2k+1}=2^{2k}\beta(2k)\,,
\end{equation}
while for even $n$:
\begin{equation}\label{I-even}
    \mathcal{I}_{2k}=-(2^{4k-2}-2^{2k-1}-2)\zeta(2k-1)\,,
\end{equation}
with $\zeta(x)$ is the Riemann zeta function and $\beta(x)$ is the Dirichlet beta function\footnote{$\mathcal I_2$ is understood as the $k\to 1$ limit of \eqref{I-even} and its value is $\mathcal I_2=-6\log 2$.}. The same special functions already appeared in the analysis of the strong coupling expansion of the cusp anomalous dimension. For example, in equations (28) and (29) of \cite{Basso:2007wd} the perturbative part of the cusp anomalous dimension were given in terms of some coefficients $c_n$ proportional to $\zeta(x)$ and $\beta(x)$ with odd and even integer arguments respectively. Hence these coefficients are in one-to-one correspondence with moments $\mathcal{I}_n$, namely:
\begin{align}\notag
    &\mathcal{I}_2=-8\pi c_1\,, && \mathcal{I}_3=256\pi^2c_2\,,\notag\\
    &\mathcal{I}_4=-4096 \pi ^3 c_3\,, && \mathcal{I}_5=\frac{262144 }{21}\pi ^4c_4\,,\notag\\
    &\mathcal{I}_6=-\frac{4194304}{87}\pi^5 c_5\,, && \mathcal{I}_7=\frac{134217728}{1605}\pi^6c_6\,,
\end{align}
etc. To avoid cumbersome expressions, in the following I will express everything in terms of $\mathcal{I}_n$.

According to \eqref{I-mod-exact} the moments $\mathcal{I}_n^{(\delta^+,\delta^-)}$ entering the higher order contributions are obtained from \eqref{I-odd} and \eqref{I-even} via:
\begin{equation}\label{Is-a=1/4}
    \mathcal{I}_n^{(\delta^+,\delta^-)}=\mathcal{I}_n-(-1)^{n-1}\sum_{l\in\delta^+}\frac{2}{\left(l+\frac{1}{4}\right)^{n-1}}+\sum_{j\in\delta^-}\frac{2}{(j+\frac{3}{4})^{n-1}}\,.
\end{equation}
Then using rule \eqref{D-rule}, all subleading $\mathcal{D}^{(\delta^+,\delta^-)}(g)$ evaluated at $a=1/4$ can be easily obtained from the perturbative part:
\begin{equation}
    \left.\mathcal D^{(\delta^+,\delta^-)}(g)\right|_{a=1/4}=\left.\mathcal{D}\left[\mathcal{I}_n^{(\delta^+,\delta^-)}\right](g)\right|_{a=1/4-\Delta}\,,
\end{equation}
where $\mathcal{D}\left[{\mathcal I_n}\right](g)$ coincides with the perturbative function in \eqref{Z00}, and the integrals $\mathcal{I}_n^{(\delta^+,\delta^-)}$ are given by \eqref{Is-a=1/4}, with $\mathcal{I}_n$ expressed in \eqref{I-odd} and \eqref{I-even}.

Substituting these integrals into the perturbative function \eqref{Z00}, evaluating them at $a=1/4-\Delta$ with the appropriate shift, and summing them up according to \eqref{degen-cusp}, arbitrary non-perturbative correction can be determined to \eqref{a=1/4-exp}.

Now I turn to the special cases of $\ell=0$ and $\ell=1$.

\subsubsection*{$\ell=0$}

For $a=1/4$ and $\ell=0$, first we have to determine the prefactor $A(g)$. By \eqref{B-guess} and equation (5.18) of \cite{Bajnok20252}, in this special case its value is given as:
\begin{equation}
    A_{\ell=0}(g)=\frac{e^{\frac{3\pi g}{4}}}{g^{\frac{5}{16}}}\Gamma\left(\frac{1}{4}\right)^{\frac{1}{4}}\Gamma\left(\frac{3}{4}\right)^{-\frac{1}{4}}\frac{e^{\frac{1}{8}}}{2^{\frac{37}{48}}\pi^{\frac{5}{16}}A^{\frac{3}{2}}}\,,
\end{equation}
Where $A$ is the Glaisher–Kinkelin constant.

In this case, by \eqref{f-nonP}, \eqref{Cs} and \eqref{Is-a=1/4}, the first few non-trivial exponentially small contributions to $Z_{\ell=0}$, up to the first three leading order terms in $1/g$, are:
\begin{align}\notag\label{C-ell0}
    \mathcal{C}^{(0)}_{\ell=0}&=1-\frac{\mathcal{I}_2}{128 \pi  g}+\frac{17 \mathcal{I}_2^2+8 \mathcal{I}_3}{32768 \pi ^2 g^2}-\frac{187 \mathcal{I}_2^3+264 \mathcal{I}_2 \mathcal{I}_3 +224 \mathcal{I}_4}{4194304 \pi ^3 g^3}+\mathcal{O}\left(\frac{1}{g^4}\right)\,,\notag\\
    \mathcal{C}^{(1)}_{\ell=0}&=-\frac{e^{\frac{i\pi}{4}}}{(8\pi g)^{\frac{1}{2}}}\frac{2\Gamma\left(\frac{3}{4}\right)}{\Gamma\left(\frac{1}{4}\right)}\left[1-\frac{9 (\mathcal{I}_2+8)}{128 \pi  g}+\frac{225 (\mathcal{I}_2+8)^2-216 (\mathcal{I}_3-32)}{32768 \pi ^2 g^2}-\right.\notag\\
    &\left.-\frac{3075 (\mathcal{I}_2+8)^3-8856 (\mathcal{I}_2+8)(\mathcal{I}_3-32)+5856 (\mathcal{I}_3+128)}{4194304 \pi ^3 g^3}+\mathcal{O}\left(\frac{1}{g^4}\right)\right]\,,\notag\\
    \mathcal C^{(3)}_{\ell=0}&=\frac{e^{-\frac{i\pi}{4}}}{(8\pi g)^{\frac{3}{2}}}\frac{2 \Gamma \left(\frac{5}{4}\right)}{3 \sqrt{3} \Gamma \left(\frac{3}{4}\right)}\left[1-\frac{25 (3 \mathcal{I}_2+8)}{384 \pi  g}+\frac{1025 (3 \mathcal{I}_2+8)^2+1000 (9 \mathcal{I}_3+32)}{294912 \pi ^2 g^2}-\right.\notag\\
    &\left.-\frac{19475 (3 \mathcal{I}_2+8)^3+57000 (3 \mathcal{I}_2+8)(9 \mathcal{I}_3+32)+37600 (27 \mathcal{I}_4+128)}{113246208 \pi ^3 g^3}+\mathcal{O}\left(\frac{1}{g^4}\right)\right]\,,
\end{align}
while $\mathcal C^{(2)}_{\ell=0}=0$. By generating higher order corrections in $1/g$ for the perturbative function $\mathcal{D}^{(\{\},\{\})}(g)$, the procedure of computing exponentially suppressed corrections to \eqref{a=1/4-exp} can be done up to arbitrary order in $e^{-2\pi n g}$.

\subsubsection*{$\ell=1$}

For $a=1/4$ and $\ell=1$, one finds for the value of the overall prefactor $A(g)$ :
\begin{equation}
    A_{\ell=1}(g)=\frac{e^{\frac{3\pi g}{4}}}{g^{\frac{21}{16}}}\Gamma\left(\frac{1}{4}\right)^{\frac{5}{4}}\Gamma\left(\frac{3}{4}\right)^{\frac{3}{4}}\frac{e^{\frac{1}{8}}}{2^{\frac{109}{48}}\pi^{\frac{21}{16}}A^{\frac{3}{2}}}\,.
\end{equation}
Repeating the same argument as for the $\ell=0$ case, the first few non-perturbative functions $\mathcal{C}^{(n)}(g)$ are:
\begin{align}\notag\label{C-ell1}
    \mathcal{C}^{(0)}_{\ell=1}&=1+\frac{15 \mathcal{I}_2}{128 \pi  g}-\frac{15\mathcal{I}_2^2+120 \mathcal{I}_3}{32768 \pi ^2 g^2}+\frac{85 \mathcal{I}_2^3+2040 \mathcal{I}_3\mathcal{I}_2+800 \mathcal{I}_4}{4194304 \pi ^3 g^3}+\mathcal{O}\left(\frac{1}{g^4}\right)\,,\notag\\
    \mathcal{C}^{(1)}_{\ell=1}&=\frac{e^{\frac{i\pi}{4}}}{(8\pi g)^{\frac{1}{2}}}\frac{2\Gamma\left(\frac{7}{4}\right)}{\Gamma\left(\frac{5}{4}\right)}\left[1+\frac{7 (\mathcal{I}_2+8)}{128 \pi  g}-\frac{63 (\mathcal{I}_2+8)^2-168 (\mathcal{I}_3-32)}{32768 \pi ^2 g^2}+\right.\notag\\
    &\left.+\frac{1575 (\mathcal{I}_2+8)^3-12600 (\mathcal{I}_2+8)(\mathcal{I}_3-32)+10080 (\mathcal{I}_4+128)}{12582912 \pi ^3 g^3}+\mathcal{O}\left(\frac{1}{g^4}\right)\right]\,,\notag\\
    \mathcal C^{(3)}_{\ell=1}&=\frac{e^{-\frac{i\pi}{4}}}{(8\pi g)^{\frac{3}{2}}}\frac{2 \Gamma \left(\frac{9}{4}\right)}{ \sqrt{3} \Gamma \left(\frac{7}{4}\right)}\left[1-\frac{9 \mathcal{I}_2+24}{128 \pi  g}+\frac{25 (9 \mathcal{I}_2+24)^2+360 (9 \mathcal{I}_3+32)}{294912 \pi ^2 g^2}-\right.\notag\\
    &\left.-\frac{1025 (9 \mathcal{I}_2+24)^3+44280 (9 \mathcal{I}_2+24) (9 \mathcal{I}_3+32) +108000 (27 \mathcal{I}_4+128)}{1019215872 \pi ^3 g^3}+\mathcal{O}\left(\frac{1}{g^4}\right)\right]\,.
\end{align}
While $\mathcal{C}^{(2)}_{\ell=1}(g)$ is again equal to zero.

\subsubsection*{Cusp anomalous dimension}

Finally, using \eqref{tilted-cusp}, I present some analytic results for the strong coupling expansion of the cusp anomalous dimension. 

If we put the expansion \eqref{a=1/4-exp} into \eqref{tilted-cusp}, we find that the strong coupling expansion of $\Gamma_{\mathrm{cusp}}$ is governed by the powers of $e^{-2\pi g}$. Hence $\Gamma_{\mathrm{cusp}}(g)$ can be parametrized as: 
\begin{equation}\label{cusp-series}
    \Gamma_{\mathrm{cusp}}(g)=\sum_{n\geq 0}e^{-2\pi n g}\gamma^{(n)}(g)\,,
\end{equation}
in agreement with \cite{Basso:2009gh}.

Substituting \eqref{C-ell0} and \eqref{C-ell1} and the values of $A(g)$ into $Z_{\ell=0}(g)$ and $Z_{\ell=1}(g)$ and expanding the ratio \eqref{tilted-cusp} in powers of $e^{-2\pi g}$ and $1/g$, for the first few $\gamma^{(n)}(g)$ corrections one obtains:
\begin{align}\notag\label{gammas}
    \gamma^{(0)}(g)&=2g\left[1+\frac{\mathcal{I}_2}{8 \pi  g}-\frac{\mathcal{I}_3}{256 \pi ^2 g^2}+\frac{2 \mathcal{I}_2 \mathcal{I}_3+\mathcal{I}_4}{4096 \pi ^3 g^3}+\mathcal{O}\left(\frac{1}{g^4}\right)\right]\,,\notag\\
    \gamma^{(1)}(g)&=e^{\frac{i\pi}{4}}\frac{2^{3 \over 2}3^{1 \over 8}g^{1 \over 2}}{\pi^{1 \over 2}k^{1\over 4}}\left[1+\frac{3+\mathcal{I}_2}{16 \pi  g}-\frac{54+12 \mathcal{I}_2+2 \mathcal{I}_2^2-\mathcal{I}_3}{1024 \pi ^2 g^2}+\right.\notag\\
    &\left.+\frac{2 \mathcal{I}_2^3+18 \mathcal{I}_2^2-3 \mathcal{I}_3 \mathcal{I}_2+162 \mathcal{I}_2-21 \mathcal{I}_3+6 \mathcal{I}_4+714}{16384 \pi ^3 g^3}+\mathcal{O}\left(\frac{1}{g^4}\right)\right]\,,\notag\\
     \gamma^{(2)}(g)&=e^{\frac{i\pi}{2}}\frac{3^{1 \over 4}}{\pi k^{1 \over 2}}\left[1-\frac{3}{8 \pi  g}+\frac{24 \mathcal{I}_2-3 \mathcal{I}_3+252}{512 \pi ^2 g^2}-\right.\notag\\
    &\left.-\frac{24 \mathcal{I}_2^2-6 \mathcal{I}_3 \mathcal{I}_2+504 \mathcal{I}_2-57 \mathcal{I}_3+4 \mathcal{I}_4+3684}{4096 \pi ^3 g^3}+\mathcal{O}\left(\frac{1}{g^4}\right)\right]\,,\notag\\
    \gamma^{(3)}(g)&=e^{-\frac{i\pi}{4} }\frac{ (k-9)}{2^{3 \over 2} 3^{13 \over 8} \pi ^{3 \over 2} k^{3 \over 4}g^{1 \over 2}}\left[1-\frac{3 (k-9) \mathcal{I}_2+5 (k-81)}{48 \pi  g (k-9)}-\right.\notag\\
    &\left.+\frac{2 (125 k-56133)+180 (k-81) \mathcal{I}_2+54 (k-9) \mathcal{I}_2^2+9 (7 k+117) \mathcal{I}_3}{9216 \pi ^2 g^2 (k-9)}+\mathcal{O}\left(\frac{1}{g^3}\right)\right]\,,
\end{align}
with the moments $\mathcal{I}_n$ given in \eqref{I-odd} and \eqref{I-even}. For convenience, I have introduced the notation:
\begin{equation}
    k=\frac{\sqrt{3} \Gamma \left(\frac{1}{4}\right)^8}{64 \pi ^4}\,.
\end{equation}
For $a=1/4$, all the Stokes constants $S^{(\delta^+,\delta^-)}$ and the prefactors $A(g)$ contain combinations of $\Gamma(n/4)$ with $|n|$ being an odd integer. Hence, using well-known relations for the Gamma functions, every function $\gamma^{(n)}(g)$ can be completely expressed in terms of the constant $k$ and the moments $\mathcal{I}_n$.

The perturbative part in \eqref{gammas} reproduces the expansion presented in \cite{Basso:2007wd}, while the non-perturbative corrections coincide with the results of \cite{Basso:2009gh} and \cite{Dorigoni:2015dha}. With the method discussed above, any higher order non-perturbative contribution to the cusp anomalous dimension can be easily obtained in an extremely efficient way. The complete strong coupling structure of $\Gamma$ is summarized in \cite{Bajnok:2026xri}.

\section{Resurgence relations}

We see that \eqref{D-rule} together with the recurrence relations \eqref{Sp-fin} and \eqref{Sm-fin} is extremely efficient to gain information about the large-$g$ expansion of the determinant with matrix Bessel kernel, hence to compute the strong coupling expansion of several physical observables in supersymmetric gauge theories. Although the method presented above is powerful to obtain analytic results, by choosing specific values of $a$ and $\ell$, higher order contributions with hundreds of $1/g$ terms can be achieved numerically up to arbitrary exponential orders.

Using relation \eqref{D-rule} and numerically evaluating $1/g$ coefficients for the functions $\mathcal{D}^{(\delta^+,\delta^-)}(g)$ with specific values of $a$ and $\ell$, one finds that these coefficients are asymptotic or, in other words, $d_k^{(\delta^+,\delta^-)}$ grow factorially for large values of $k$. This means that the transseries in \eqref{trans-D} is only formal and a resurgence analysis \citep{Marino:2012zq,Dorigoni:2014hea,Aniceto:2018bis} is essential to be able to resum the transeries and obtain a reasonable physical answer.

In \cite{Bajnok20252} we have already investigated the resurgence properties of the determinant in representation \eqref{eq:trans_gamma}. There we have studied the asymptotic behavior of the leading non-perturbative contributions with high precision numerical analysis. Although we found that certain contributions are connected via direct resurgence relations, it turned out that not all of them are related, and the reason of the absence of certain relations remained unknown. The structure of direct connections is illustrated in Figure 3. of \cite{Bajnok20252}. In this section I will follow the ideas discussed in Appendix A to derive some basic resurgence properties for the restructured strong coupling expansion \eqref{trans-D}. I show that the asymptotic behavior of the non-perturbative functions and the analytic structure of their Borel transforms immediately follow from the transseries structure and that the new form of the expansion shed light to the resurgence structure found in \cite{Bajnok20252}. With high precision numerical analysis, I verify these resurgence relations and by that, I give another strong evidence on the validity of \eqref{trans-D}.

Since in the transseries \eqref{trans-D}, the coupling constant $g$ only appears together with a factor of $8\pi$, to avoid lengthy expressions, in the following I will use the notation:
\begin{equation}
    \tilde g=8\pi g=2\sqrt{\lambda}\,.
\end{equation}
In terms of the rescaled coupling constant $\tilde g$, with a slight abuse of notation, the strong coupling expansion takes the form:
\begin{align}
    Z_\ell(g)=A_\ell( g)&\sum_{\delta^+,\delta^-}\tilde g^{-\Delta(\Delta-2a)}e^{-\tilde g \left(\sum_{l\in \delta^+}  x_l^++\sum_{j\in \delta^- } x_j^-\right)}e^{i\pi a\Delta }S^{(\delta^+,\delta^-)}\mathcal{D}^{(\delta^+,\delta^-)}(\tilde g)\,.
\end{align}
$\mathcal{D}^{(\delta^+,\delta^-)}(g)$ given by the series:
\begin{equation}\label{D-tildeg}
    \mathcal{D}^{(\delta^+,\delta^-)}(\tilde g)=\sum_{k\geq 0}\frac{d_k^{(\delta^+,\delta^-)}}{\tilde g^k}\,,
\end{equation}
with the same expansion coefficients $d_k^{(\delta^+,\delta^-)}$ as before.

\subsection{Stokes automorphism}

The asymptotic behavior of the perturbative function $\mathcal{D}^{(\{\},\{\})}$ is captured by the analytic properties of its Borel transform:
\begin{equation}\label{borel-main}
    \mathcal B\left[\mathcal{D}^{\left(\{\},\{\}\right)}\right](s)=\sum_{k\geq0}\frac{d_k^{(\{\},\{\})}s^k}{\Gamma(k+1)}\,.
\end{equation}
For a convergent series of the form \eqref{D-tildeg}, the function $\mathcal B\left[\mathcal{D}^{\left(\{\},\{\}\right)}\right](s)$ can be resummed, and its inverse, given by the integral \eqref{inv-borel} over the positive real line, gives back the original function.

However, since the coefficients $d_k^{(\{\},\{\})}$ grow factorially with $k$, the function $\mathcal B\left[\mathcal{D}^{\left(\{\},\{\}\right)}\right](s)$ produces singularities and cuts on the complex $s$ plane, called the Borel plane. These singularities should be avoided while taking the invers transformation either below, or above the real axis. Therefore, one has to define the lateral Borel resummations:
\begin{equation}
    \mathcal S_\pm\left[\mathcal D^{(\{\},\{\})}\right](\tilde g)=\tilde g\int_{0}^{\infty e^{\pm i\epsilon}} ds e^{-\tilde g s}\mathcal{B}\left[\mathcal D^{(\{\},\{\})}\right](s)\,.
\end{equation}
This means that there is an ambiguity in the resummation of $\mathcal{D}^{(\{\},\{\})}(\tilde{g})$. This ambiguity is the reason of the appearance of non-perturbative contributions in the transseries \eqref{trans-D}. The same argument holds for the higher order, exponentially small corrections as well.

The two lateral Borel resummations $\mathcal{S}_+$ and $\mathcal{S}_-$ are related by the Stokes automorphism $\mathfrak S$:
\begin{equation}
    \mathcal{S}_+=\mathcal{S}_-\circ \mathfrak S\,.
\end{equation}
Its logarithm is defined by the so called Alien derivatives $\Delta_\omega$:
\begin{equation}\label{ln-stokes-pm}
    \ln \mathfrak S=\sum_\omega e^{-\tilde g \omega}\Delta_\omega\,,
\end{equation}
where $\omega$ denotes the singular points of \eqref{borel-main} on the real line. The Alien derivatives describe the discontinouities of the Borel transform across the singular points $\omega$.

For a given transseries, such as \eqref{trans-D}, obtained from some system of integro-differential equations, the Alien derivatives can be computed via the Bridge equations \citep{Marino:2012zq,Dorigoni:2014hea,Aniceto:2018bis}. If we assign a parameter $\sigma_l^\pm$ to every distinct exponential weight $x_l^\pm$, we can define:
\begin{align}
    Z_\ell\left(g,\{\sigma^\pm\}\right)=A_\ell(\tilde g)&\sum_{\delta^+,\delta^-}\left(\prod_{l\in \delta^+}\sigma_l^+ e^{-\tilde g x_l^+}\right)\left(\prod_{j\in \delta^-}\sigma_j^-e^{-\tilde gx_j^-}\right)\tilde{g}^{-\Delta(\Delta-2a)}e^{i\pi a\Delta }S^{(\delta^+,\delta^-)}\mathcal{D}^{(\delta^+,\delta^-)}(\tilde{g})\,.
\end{align}
Then the Bridge equations connect the Alien derivatives of $Z_\ell\left(g,\{\sigma^\pm\}\right)$ to ordinary differentials with respect to the parameters $\sigma_l^\pm$. In Appendix A, using the Bridge equations, I derive the Alien derivatives for a similar, but somewhat less complicated transseries than \eqref{trans-D}. It is simpler in the sense, that it contains only one set of exponential weights $\{x_l\}_{l\geq0}$ instead of having two sectors $\{x^\pm_l\}_{l\geq0}$, and it does not contain the complex phases $e^{i\pi a\Delta}$. However, for the relevant values $0<a<1/2$, the physical weights can be ordered as $x_0^+<x_0^-<x_1^+<x_1^-<\dots$, therefore, we can assign $x_{2l}\to x^+_l$, $x_{2l+1}\to x^-_l$, and redefine $\mathcal{D}^{(\delta^+,\delta^-)}(\tilde g)$ by collecting its prefactors, so we arrive at the same transseries as in \eqref{F-toy}. The power $\lambda(\delta)$ of the leading $1/g$ terms is given by $-\Delta(\Delta-2a)$ in this case.

Repeating the same argument as in Appendix A, one finds that the singular points of $\mathcal{B}\left[\mathcal{D}^{(\delta^+,\delta^-)}\right](s)$ appear on the positive real line at $x^\pm_j$ with $j\notin \delta^+,\delta^-$ and the corresponding Alien derivatives are:
\begin{align}\notag\label{AD-pm}
    &\Delta^+_j\mathcal D^{(\delta^+,\delta^-)}(\tilde g)=\tilde g^{-1-2\Delta+2a}A_j^+ \frac{S^{(\delta^+\cup\{j\},\delta^-)}}{S^{(\delta^+,\delta^-)}}\mathcal{D}^{(\delta^+\cup\{j\},\delta^-)}(\tilde g)\,,\\
    &\Delta^-_j\mathcal D^{(\delta^+,\delta^-)}(\tilde g)= \tilde g^{-1+2\Delta-2a}A_j^-\frac{S^{(\delta^+,\delta^-\cup\{j\})}}{S^{(\delta^+,\delta^-)}}\mathcal{D}^{(\delta^+,\delta^-\cup\{j\})}(\tilde g)\,.
\end{align}
The factors $A_j^\pm$ are yet undetermined numbers, and for simplicity, I have introduced the notation $\Delta_j^\pm\equiv \Delta_{x^\pm_j}$. For $j\in\delta^+$ or $j\in \delta^-$ the Alien derivatives give zero.

From the Bridge equations it also follows that \eqref{AD-pm} is supplemented with the relations:
\begin{equation}\label{AD-pm-algebra}
    \left(\Delta^\pm_j\right)^2=0,\qquad \left[\Delta^+_l,\Delta^+_j\right]=\left[\Delta^-_l,\Delta^-_j\right]=\left[\Delta^+_l,\Delta^-_j\right]=0\,,
\end{equation}
for all $j,l\geq 0$ integers.

Equations \eqref{AD-pm} and \eqref{AD-pm-algebra} mean that $\mathcal{D}^{(\delta^+,\delta^-)}(\tilde{g})$ resurges only with the subleading corrections $\mathcal{D}^{(\delta^+\cup\{j\},\delta^-)}(\tilde{g})$ and $\mathcal{D}^{(\delta^+,\delta^-\cup\{j\})}(\tilde{g})$ (so that their exponential orders differ only by a factor of $e^{-\tilde g x^\pm_j}$). This is in accordance with the findings of \cite{Bajnok20252}. There we found a direct resurgence relation, for example, between the $(n,m)=(0,0)$ and $(n,m)=(2,1)$ terms of \eqref{eq:trans_gamma}, but not between the $(n,m)=(0,0)$ and $(n,m)=(2,2)$ contributions. The $(n,m)=(0,0)$ perturbative part belongs to $(\delta^+,\delta^-)=(\{\},\{\})$ in the \eqref{trans-D} prescription. Due to \eqref{AD-pm} it has a direct resurgence relation with $(\delta^+,\delta^-)=(\{1\},\{\})$, which by \eqref{delta-nm-eg} corresponds to $(n,m)=(2,1)$. In contrast, the $(\delta^+,\delta^-)=(\{1\},\{0\})$ term is not related to the perturbative part, since they differ in more than two different exponential factors of $e^{-\tilde g x^\pm_j}$. Therefore, by \eqref{delta-nm-eg}, $(n,m)=(0,0)$ should not be in direct resurgence relation with $(n,m)=(2,2)$. This is in complete agreement with Figure 3. of \cite{Bajnok20252}.

The Alien algebra in \eqref{AD-pm} and \eqref{AD-pm-algebra} is useful to determine the asymptotic behavior of the coefficients $d_k^{(\delta^+,\delta^-)}$. Using the algebra \eqref{AD-pm-algebra} we find from definition \eqref{ln-stokes-pm} that $\mathfrak S$ is:
\begin{equation}\label{exp-stokes-AD}
    \mathfrak S=\prod_{l\geq 0}\left(1+e^{-x_l^+ \tilde g}\Delta^+_l\right)\prod_{j\geq 0}\left(1+e^{-x_j^- \tilde g}\Delta^-_j\right)\,.
\end{equation}
As it is discussed in Appendix A, by acting with $\mathfrak S$ on the non-perturbative functions $\mathcal{D}^{(\delta^+,\delta^-)}(g)$, one can find that the difference between the lateral resummations, hence the discontinouities of their Borel transforms can be easily related to the subleading strong coupling corrections. 

\subsection{Asymptotic analysis}

In the following I illustrate how the Stokes automorphism \eqref{exp-stokes-AD} gives the opportunity to determine the asymptotic behavior of the non-perturbative corrections. For simplicity, I present the method only for a special set of corrections that correspond to labels of the form $(\delta^+,\delta^-)=(\{0,1,\dots,l\},\{0,1,\dots,l-1\})$ and $(\delta^+,\delta^-)=(\{0,1,\dots,l\},\{0,1,\dots,l\})$, where the elements of $\delta^+$ and $\delta^-$ go along all non-negative integers up to $l$ and $l-1$. These corrections are special, since by numerically investigating the asymptotic behavior of their $1/g$ expansions, the coefficients $A_l^\pm$ can be determined and the results for the corresponding Stokes constants $S^{(\delta^+,\delta^-)}$ can be tested.

Since the exponential weights are ordered as $x_0^+<x_0^-<x_1^+<x_1^-<\dots$, by acting with $\mathfrak{S}$ on the first few such corrections and using \eqref{AD-pm} we find:
\begin{align}\notag\label{disc-spec}
    &\mathcal{S}_+\left[\mathcal{D}^{(\{\},\{\})}\right](\tilde g)-\mathcal{S}_-\left[\mathcal{D}^{(\{\},\{\})}\right](\tilde g)=\tilde g^{-1+2a}e^{-\tilde gx_0^+}A^+_{0}S^{(\{0\},\{\})}\mathcal{D}^{(\{0\},\{\})}(\tilde g)+\dots\,,\notag\\
    &\mathcal{S}_+\left[\mathcal{D}^{(\{0\},\{\})}\right](\tilde g)-\mathcal{S}_-\left[\mathcal{D}^{(\{0\},\{\})}\right](\tilde g)=\tilde g^{1-2a}e^{-\tilde gx_0^-}A^-_{0}\frac{S^{(\{0\},\{0\})}}{S^{(\{0\},\{\})}}\mathcal{D}^{(\{0\},\{0\})}(\tilde g)+\dots\,,\notag\\
    &\mathcal{S}_+\left[\mathcal{D}^{(\{0\},\{0\})}\right](\tilde g)-\mathcal{S}_-\left[\mathcal{D}^{(\{0\},\{0\})}\right](\tilde g)=\tilde g^{-1+2a}e^{-\tilde gx_1^+}A^+_{1}\frac{S^{(\{0,1\},\{0\})}}{S^{(\{0\},\{0\})}}\mathcal{D}^{(\{0,1\},\{0\})}(\tilde g)+\dots\,,\notag\\
    &\mathcal{S}_+\left[\mathcal{D}^{(\{0,1\},\{0\})}\right](\tilde g)-\mathcal{S}_-\left[\mathcal{D}^{(\{0,1\},\{0\})}\right](\tilde g)=\tilde g^{1-2a}e^{-\tilde gx_1^-}A^-_{1}\frac{S^{(\{0,1\},\{0,1\})}}{S^{(\{0,1\},\{0\})}}\mathcal{D}^{(\{0,1\},\{0,1\})}(\tilde g)+\dots\,,
\end{align}
and so on. The dots denote exponentially suppressed terms. Then the leading asymptotic expansion of the $1/g$-coefficients can be parametrized as:
\begin{equation}\label{d-asymp}
    d_k^{(\delta^+,\delta^-)}=\frac{1}{\pi}\sum_{j\geq 0}c_j\frac{\Gamma(k+\lambda-j)}{A^{k+\lambda-j}}+\dots\,,
\end{equation}
where the parameters $A$, $\lambda$ and $c_j$ can be fixed by \eqref{disc-spec}. By taking the Borel transform \eqref{borel-main} of $\mathcal{D}^{(\{\},\{\})}(\tilde g)$, $\mathcal{D}^{(\{0\},\{\})}(\tilde g)$, etc. parametrized as \eqref{d-asymp}, and evaluating their lateral Borel resummations $\mathcal{S}_\pm$, their difference should be compared with \eqref{disc-spec}. From the jumps across the leading cuts, one finds:
\begin{align}\notag\label{asym-d-spec}
    &d_k^{(\{\},\{\})}=\frac{A_0^+}{2\pi i}S^{(\{0\},\{\})}\sum_{j\geq 0}d_k^{(\{0\},\{\})}\frac{\Gamma(k-j-1+2a)}{(x_0^+)^{k-j-1+2a}}+\dots\,,\notag\\
    &d_k^{(\{0\},\{\})}=\frac{A_0^-}{2\pi i}\frac{S^{(\{0\},\{0\})}}{S^{(\{0\},\{\})}}\sum_{j\geq 0}d_k^{(\{0\},\{0\})}\frac{\Gamma(k-j+1-2a)}{(x_0^-)^{k-j+1-2a}}+\dots\,,\notag\\
    &d_k^{(\{0\},\{0\})}=\frac{A_1^+}{2\pi i}\frac{S^{(\{0,1\},\{0\})}}{S^{(\{0\},\{0\})}}\sum_{j\geq 0}d_k^{(\{0,1\},\{0\})}\frac{\Gamma(k-j-1+2a)}{(x_1^+)^{k-j-1+2a}}+\dots\,,\notag\\
    &d_k^{(\{0,1\},\{0\})}=\frac{A_1^-}{2\pi i}\frac{S^{(\{0,1\},\{0,1\})}}{S^{(\{0,1\},\{0\})}}\sum_{j\geq 0}d_k^{(\{0,1\},\{0,1\})}\frac{\Gamma(k-j+1-2a)}{(x_1^-)^{k-j+1-2a}}+\dots\,,
\end{align}
where the dots denote corrections that correspond to subleading cuts. The leading asymptotic behavior of higher order contributions with $(\delta^+,\delta^-)=(\{0,1,\dots,l\},\{0,1,\dots,l-1\})$ and $(\delta^+,\delta^-)=(\{0,1,\dots,l\},\{0,1,\dots,l\})$ contain the overall prefactors $A^-_l$ and $A^+_l$ respectively. Since with formula \eqref{D-rule} the coefficients $d_k^{(\delta^+,\delta^-)}$ can be determined numerically for arbitrary $\delta^\pm$ efficiently, the asymptotic behaviors in \eqref{asym-d-spec} could be verified and the coefficients $A^\pm_l$ could be extracted.

For the parameters $(a,\ell)=(1/4,0)$, $(a,\ell)=(1/4,1)$ and $(a,\ell)=(1/(2\sqrt{2}),2)$ I have generated $d_k^{(\delta^+,\delta^-)}$ for $(\delta^+,\delta^-)=(\{\},\{\}),(\{0\},\{\}),(\{0\},\{0\}),(\{0,1\},\{0\}),(\{0,1\},\{0,1\})$ and $(\{0,1,2\},\{0,1\})$ up to $k=200$ with $500$ digit precision. The values $(a,\ell)=(1/4,0)$, $(a,\ell)=(1/4,1)$ for the parameters $a$ and $\ell$ were motivated by physical relevance (see the discussions above about the cusp anomalous dimension), while the value $a=1/(2\sqrt{2})$ was chosen to be an irrational number in the range $0<a<1/2$ to have a clear separation between the non-perturbative contributions.

First, I computed the Borel transforms of the above-mentioned non-perturbative series and took their diagonal Padé approximations. With a suitable conformal map (see Appendix G of \cite{Bajnok20252}) I separated their different cuts in the Borel plane. The analytical structures of the Borel transforms of $\mathcal D^{(\{0\},\{\})}(\tilde g)$, $\mathcal D^{(\{0\},\{0\})}(\tilde g)$ and $\mathcal D^{(\{0,1\},\{0\})}(\tilde g)$ are depicted in Figure 2. The cut structure is in agreement with the argument given in Section 3. The Borel transform of the perturbative function $\mathcal{D}^{(\{\},\{\})}(\tilde g)$ has cuts on the negative real line of the Borel plane starting from $-y_i$ and their linear combinations, while on the positive real line starting from the combinations of $x_l^\pm$. They correspond to the poles and zeros of $\Phi_\pm(x)$, respectively. To obtain the non-perturbative contribution $\mathcal{D}^{(\{0\},\{\})}(\tilde g)$ from the perturbative one, one has to promote the first zero $x_0^+$ of $\Phi_+(x)$ to be a pole instead. This means that the cuts on the positive real line starting from values, which contain $x_0^+$, are removed and new cuts appear on the negative side starting from $-x_0^+$ and its linear combinations with $-y_i$. In case of going from $\mathcal{D}^{(\{0\},\{\})}(\tilde g)$ to $\mathcal{D}^{(\{0\},\{0\})}(\tilde g)$, the same thing happens with the cut starting from $x_0^-$, and so on. This phenomenon is reflected in the cut structures illustrated in Figure 2.
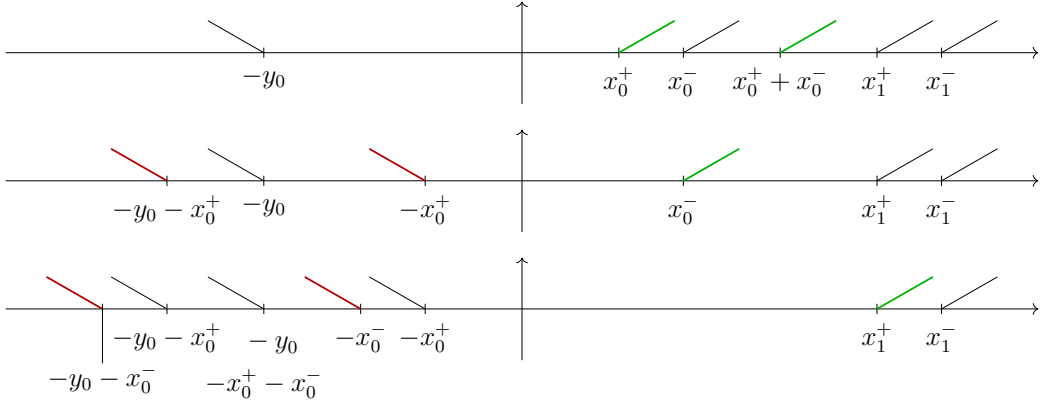
\begin{figure}[t]
  \centering
  \resizebox{0.8\linewidth}{!}{\begin{tikzpicture}

  \begin{scope}[yshift=0cm]
    \draw[->] (-8,0) -- (8,0);
    \draw[->] (0,-0.8) -- (0,0.8);

    \foreach \x/\lab in {
      -4/{-y_0},
      1.5/{x_0^+},
      2.5/{x_0^-},
      4/{x_0^+ +x_0^-},
      5.5/{x_1^+},
      6.5/{x_1^-}
    } {
      \draw (\x,0.08) -- (\x,-0.08);
      \node[below] at (\x,-0.08) {$\lab$};

      \ifdim \x pt < 0pt
        \draw (\x,0) -- ++(150:1);
      \else
        \ifdim \x pt = 1.5pt
          \draw[green!70!black, thick] (\x,0) -- ++(30:1);
        
        \else

        \ifdim \x pt = 4pt
          \draw[green!70!black, thick] (\x,0) -- ++(30:1);

          \else
        
          \draw (\x,0) -- ++(30:1);

          \fi
        \fi
      \fi
    }
  \end{scope}

  \begin{scope}[yshift=-2cm]
    \draw[->] (-8,0) -- (8,0);
    \draw[->] (0,-0.8) -- (0,0.8);

    \foreach \x/\lab in {
      -5.5/{-y_0-x_0^+},
      -4/{-y_0},
      -1.5/{-x_0^+},
      2.5/{x_0^-},
      5.5/{x_1^+},
      6.5/{x_1^-}
    } {
      \draw (\x,0.08) -- (\x,-0.08);
      \node[below] at (\x,-0.08) {$\lab$};

      \ifdim \x pt < 0pt
      \ifdim \x pt = -1.5pt
        \draw[red!70!black, thick] (\x,0) -- ++(150:1);
      \else 
      \ifdim \x pt = -5.5pt
        \draw[red!70!black, thick] (\x,0) -- ++(150:1);
        \else
        \draw (\x,0) -- ++(150:1);
        \fi
        \fi
      \else
        \ifdim \x pt = 2.5pt
          \draw[green!70!black, thick] (\x,0) -- ++(30:1);
        \else
          \draw (\x,0) -- ++(30:1);
        \fi
      \fi
    }
  \end{scope}

  \begin{scope}[yshift=-4cm]
    \draw[->] (-8,0) -- (8,0);
    \draw[->] (0,-0.8) -- (0,0.8);

    \foreach \x/\lab in {
      -6.5/{-y_0-x_0^-},
      -5.5/{-y_0-x_0^+},
      -4/{\begin{aligned}
          &-y_0\\
          -x&_0^+-x_0^-
      \end{aligned}},
      -2.5/{-x_0^-},
      -1.5/{-x_0^+},
      5.5/{x_1^+},
      6.5/{x_1^-}
    } {
    \ifdim \x pt = -6.5 pt
      \draw (\x,0.08) -- (\x,-0.85);
    \else
      \draw (\x,0.08) -- (\x,-0.08);
    \fi
      
      \ifdim \x pt = -6.5 pt
      \node[below, yshift=-18pt] at (\x,-0.08) {$\lab$};
    \else
      \node[below] at (\x,-0.08) {$\lab$};
    \fi

      \ifdim \x pt < 0pt
        \ifdim \x pt = -2.5pt
        \draw[red!70!black, thick] (\x,0) -- ++(150:1);
      \else 
      \ifdim \x pt = -6.5pt
        \draw[red!70!black, thick] (\x,0) -- ++(150:1);
        \else
        \draw (\x,0) -- ++(150:1);
        \fi
        \fi
      \else
        \ifdim \x pt = 5.5pt
          \draw[green!70!black, thick] (\x,0) -- ++(30:1);
        \else
          \draw (\x,0) -- ++(30:1);
        \fi
      \fi
    }
  \end{scope}

\end{tikzpicture}%
}
  \caption{The figures from top to bottom illustrate the cut structures of $\mathcal{B}\left[\mathcal D^{(\delta^+,\delta^-)}\right](s)$ on the Borel plane with $(\delta^+,\delta^-)=(\{\},\{\}),(\{0\},\{\})$ and $(\{0\},\{0\})$. The black lines represent cuts starting from the branch points on the real line and ending at infinity. The green lines represent which cuts has to be removed from the positive line to go from one correction to the next one. The red lines appearing on the negative real line denote the additional cuts compared to the previous correction.}
  \label{fig:imported-tikz}
\end{figure}

As a next step, with the method discussed in Appendix G of \cite{Bajnok20252}, I was able to capture the asymptotic behaviors of $d_k^{(\{0\},\{\})}$, $d_k^{(\delta^+,\delta^-)}$ with $(\delta^+,\delta^-)=(\{\},\{\}),(\{0\},\{\}),(\{0\},\{0\}),(\{0,1\},\{0\})$ and $(\{0,1\},\{0,1\})$ around the leading cuts. I was able to verify that with some suitable values of $A_j^\pm$, the asymptotic expansions in \eqref{asym-d-spec} hold. 

For each value of $(a,\ell)$ mentioned above, with a $10^{-6}$ relative error, I found that the first few $A_l^\pm$ are the following:
\begin{align}\notag\label{A-exam}
    &A_0^+=-2i\sin\left(\pi a\right)\,, &&A_0^-=2i\sin\left(\pi a\right)\,, \notag\\
    &A_1^+=-2i\sin\left(\pi a\right)\,, && A_1^-=2i\sin\left(\pi a\right)\,, \notag\\
    &A_2^+=-2i\sin\left(\pi a\right)\,. 
\end{align}
Notice that the leading asymptotic behavior of $d^{(\{\},\{\})}$, $d^{(\{0\},\{\})}$ and $d^{(\{\},\{0\})}$ with the coefficients given in \eqref{A-exam} is compatible with results found in \cite{Bajnok20252} for the functions $\mathcal Z^{(0,0)}(g)$, $\mathcal Z^{(1,0)}(g)$ and $\mathcal Z^{(0,1)}(g)$, however there are sign differences in the coefficients $A_0^+$ and $A_0^-$. As a verification of \eqref{A-exam}, I also performed a Richardson extrapolation on the coefficients of the perturbative function. Its asymptotic behavior agreed with \eqref{asym-d-spec} with the same coefficient $A_0^+$ as in \eqref{A-exam}. Furthermore, with the method discussed in Appendix G of \cite{bajnok20253}, I was also able to separate the first subleading cut of $\mathcal{B}\left[\mathcal D^{(\{\},\{\})}\right](s)$. I found that around this cut, its asymptotic behavior is given by:
\begin{equation}
    d_k^{(\{\},\{\})}\sim\frac{A_0^-}{2\pi i}S^{(\{\},\{0\})}\sum_{j\geq 0}d_k^{(\{\},\{0\})}\frac{\Gamma(k-j-1+2a)}{(x_0^-)^{k-j-1+2a}}\,,
\end{equation}
with $A_0^-$ being the same as in \eqref{A-exam} also in agreement with \eqref{AD-pm}.

The values found in \eqref{A-exam} suggest that the general form of the coefficients $A_l^\pm$ is given by:
\begin{equation}\label{A-gen}
    A^\pm_l=\mp 2i\sin\left(\pi a\right)\,.
\end{equation}
Notice that, as expected, $A^+_l$ is related to $A^-_l$ via $a\to -a$. With this the Alien derivatives in \eqref{AD-pm} become:
\begin{align}\notag\label{Alien-fin}
    &\Delta^+_j\mathcal D^{(\delta^+,\delta^-)}(\tilde g)=-2i\sin\left(\pi a\right)\,\tilde g^{-1-2\Delta+2a}\frac{S^{(\delta^+\cup\{j\},\delta^-)}}{S^{(\delta^+,\delta^-)}}\mathcal{D}^{(\delta^+\cup\{j\},\delta^-)}(\tilde g)\,,\\
    &\Delta^-_j\mathcal D^{(\delta^+,\delta^-)}(\tilde g)= 2i\sin\left(\pi a\right)\,\tilde g^{-1+2\Delta-2a}\frac{S^{(\delta^+,\delta^-\cup\{j\})}}{S^{(\delta^+,\delta^-)}}\mathcal{D}^{(\delta^+,\delta^-\cup\{j\})}(\tilde g)\,.
\end{align}

\subsection{Median resummation}

The appearance of trigonometric factors $\sin(\pi a)$ in the Alien derivatives is due to the complex phases $e^{i\pi a\Delta}$ in the transseries. As discussed in \cite{Bajnok20252}, strictly speaking, the strong coupling expansion in \eqref{eq:trans_gamma} is only valid for $\Im g<0$ and corresponds to the lateral resummation $\mathcal{S}_+$. The same holds for the transseries \eqref{trans-D}, namely the resummation of $Z_\ell(g)$ is understood as:
\begin{align}\label{trans-D-Sp}
    Z_\ell(g)=A_\ell( g)&\sum_{\delta^+,\delta^-}\tilde{g}^{-\Delta(\Delta-2a)}e^{-\tilde g \left(\sum_{l\in \delta^+}  x_l^++\sum_{j\in \delta^- } x_j^-\right)}e^{i\pi a\Delta }S^{(\delta^+,\delta^-)}\mathcal{S}_+\left[\mathcal{D}^{(\delta^+,\delta^-)}\right](\tilde{g})\,.
\end{align}
If we use $\mathcal S_-$ instead, the complex conjugate of \eqref{trans-D-Sp} should be taken:
\begin{align}\label{trans-D-Sm}
    Z_\ell(g)=A_\ell( g)&\sum_{\delta^+,\delta^-}\tilde{g}^{-\Delta(\Delta-2a)}e^{-\tilde g \left(\sum_{l\in \delta^+}  x_l^++\sum_{j\in \delta^- } x_j^-\right)}e^{-i\pi a\Delta }S^{(\delta^+,\delta^-)}\mathcal{S}_-\left[\mathcal{D}^{(\delta^+,\delta^-)}\right](\tilde{g})\,,
\end{align}
which is valid for $\Im g>0$.

If we consider, for instance, the perturbative part, the ambiguity in choosing either $\mathcal{S}_+\left[\mathcal{D}^{(\{\},\{\})}\right](\tilde{g})$ or $\mathcal{S}_-\left[\mathcal{D}^{(\{\},\{\})}\right](\tilde{g})$ should cancel the $(\{0\},\{\})$ term. This means that:
\begin{equation}
    \mathcal{S}_+\left[\mathcal{D}^{(\{\},\{\})}\right](\tilde{g})-\mathcal{S}_-\left[\mathcal{D}^{(\{\},\{\})}\right](\tilde{g})\sim-(e^{i\pi a}-e^{-i\pi a})\tilde g^{-1+2a}e^{-\tilde g x_0^+}S^{(\{0\},\{\})}\mathcal{D}^{(\{0\},\{\})}(\tilde g)\,.
\end{equation}
The prefactor on the right-hand side is exactly the constant $A_0^+$ in \eqref{A-exam}. Similarly, the discontinuity around the first subleading cut cancels against:
\begin{equation}
-(e^{-i\pi a}-e^{i\pi a})\tilde g^{-1-2a}e^{-\tilde g x_0^+}S^{(\{\},\{0\})}\mathcal{D}^{(\{\},\{0\})}(\tilde g)\,,    
\end{equation}
which gives $A_0^-=2i\sin\left(\pi a\right)$. The same argument holds for higher order contributions in agreement with \eqref{A-gen}.

As we see, neither \eqref{trans-D-Sp} nor \eqref{trans-D-Sm} are real functions of the coupling constant; therefore, they do not give a proper physical answer for the strong coupling expansion of the observables given by the determinant \eqref{F-def}. Instead, the physical result is obtained via the median resummation. 

The median resummation is defined as:
\begin{equation}
    \mathcal S_{\mathrm{med}}=\mathcal{S}_+\circ \mathfrak S^{-1/2}= \mathcal{S}_-\circ \mathfrak S^{1/2}\,.
\end{equation}
The square root of the Stokes automorphism can be easily expressed in terms of Alien derivatives by expanding the square root of \eqref{exp-stokes-AD} and using the algebraic properties in \eqref{AD-pm}. Then by \eqref{Alien-fin}, the effect of $\mathfrak S^{\pm 1/2}$ on the non-perturbative functions $\mathcal{D}^{(\delta^+,\delta^-)}(\tilde g)$ can be easily determined.

After substituting $\mathcal S_+=\mathcal S_{\mathrm{med}}\circ \mathfrak S^{1/2}$ into \eqref{trans-D-Sp}, and, for convenience, redefining the non-perturbative functions as:
\begin{equation}
    \bar{\mathcal D}^{(\delta^+,\delta^-)}(\tilde g)\equiv \tilde{g}^{-\Delta(\Delta-2a)}\frac{S^{(\delta^+,\delta^-)}}{\left(\prod_{l\in\delta^+}S^{\{l\}}_+\right)\left(\prod_{j\in\delta^-}S^{\{j\}}_-\right)}\mathcal D^{(\delta^+,\delta^-)}(\tilde g)\,,
\end{equation}
with $S_\pm^{\{l\}}$ given in \eqref{Selem}, we find that the median resummation of the determinant $Z_\ell(g)$ is given by:
\begin{equation}\label{Z-median}
    Z_\ell\left(g\right)=A_\ell(g)\sum_{\delta^+,\delta^-}\left(\prod_{l\in \delta^+}\sigma_l^+ e^{-\tilde g x_l^+}\right)\left(\prod_{j\in \delta^-}\sigma_j^-e^{-\tilde gx_j^-}\right)\mathcal{S}_{\mathrm{med}}\bar {\mathcal{D}}^{(\delta^+,\delta^-)}(\tilde{g})\,,
\end{equation}
with the parameters $\sigma_l^\pm$ being:
\begin{equation}
    \sigma^\pm_l=\cos\left(\pi a\right) S_\pm^{\{l\}}\,.
\end{equation}
Notice that \eqref{Z-median} is now a real function of the coupling constant $\tilde g$. Therefore, the median resummation in \eqref{Z-median} provides the physical answer for the strong coupling expansion of the determinant with matrix Bessel kernel.

This completes the full resurgence structure of the strong coupling expansion of \eqref{F-def} and provides strong evidence on the validity of the transseries structure \eqref{trans-D}. As a physical example, in \cite{Bajnok:2026xri} we summarize the complete strong coupling and resurgence structure of the cusp anomalous dimension.

\section*{Acknowledgements}

I would like to thank Zoltan Bajnok, Gregory P. Korchemsky and Dennis Le Plat for useful discussions and helpful comments on the draft of this paper. The research was supported by the Doctoral Excellence Fellowship Programme
funded by the National Research Development and Innovation Fund of
the Ministry of Culture and Innovation and the Budapest University
of Technology and Economics, under a grant agreement with the National
Research, Development and Innovation Office (NKFIH). It was also supported by the grant NKKP Advanced 152467.

\appendix

\section{Bridge equations}

In this appendix, based on \citep{Marino:2012zq,Dorigoni:2014hea,Aniceto:2018bis} I derive the Alien algebra for a transseries of the form:
\begin{equation}\label{F-toy}
    F(g)=\sum_{\delta}e^{-g \sum_{l\in \delta}  x_l }g^{\lambda(\delta)}\Phi^{(\delta)}(g)\,,
\end{equation}
with an infinite set of distinct exponential weights $x_0<x_1<\dots$. The summation goes over all possible unordered sets $\delta$ of distinct non-negative integers. This means that the sum contains terms that are at most first order in each factor of $e^{-gx_l}$. The functions $\Phi^{(\delta)}$ are given by the expansions:
\begin{equation}\label{Phi-series}
    \Phi^{(\delta)}(g)=\sum_{k\geq0}\frac{\phi_k^{(\delta)}}{g^k}\,,
\end{equation}
and $\lambda(\delta)$ is a function of $\delta$ that governs the leading $1/g$ order at exponential level $\delta$. For simplicity in this Appendix I considered only one sector of exponential weights. Although the large-$g$ expansion of the determinant $Z(g)$ in \eqref{trans-D} contains two sets of weights $x^+_j$ and $x^-_j$, the argument below remains the same for $Z(g)$ as well. The function $F(g)$ with $\lambda(\delta)=0$ and $g\to8\pi g$ coincides with the strong coupling expansions investigated in \cite{bajnok20253}.

The non-perturbative structure is deeply connected to the large order behavior of the coefficients $\phi_k^{(\delta)}$. This property is captured by the Borel transformation:
\begin{equation}\label{phi-borel} \mathcal{B}\left[\Phi^{(\delta)}\right](s)=\sum_{k\geq 0}\frac{\phi_k^{(\delta)}}{\Gamma(k+1)}s^k\,.
\end{equation}
If the series \eqref{Phi-series} is convergent, then its inverse is simply a Laplace transform:
\begin{equation}\label{inv-borel}
    \mathcal S\left[\Phi^{(\delta)}\right](g)=g\int_{0}^\infty ds e^{-gs}\mathcal{B}\left[\Phi^{(\delta)}\right](s)\,.
\end{equation}
and it gives back the original function $\Phi^{(\delta)}(g)$. 

The problem arises if the coefficients $\phi_k^{(\delta)}$ grow factorially, for example if they have the asymptotic form:
\begin{equation}\label{phi-asymp}
    \phi_k^{(\delta)}=\frac{1}{\pi}\sum_{j\geq 0}c_j\frac{\Gamma(k+\lambda-j)}{A^{k+\lambda-j}}\,,
\end{equation}
where $A$, $\lambda$ and $c_j$ (with $j\geq0$) are different parameters of the expansion. In general, the expansion of $\phi_k^{(\delta)}$ is a linear combination of terms similar to \eqref{phi-asymp} with various sets of parameters.

In this case, the series in \eqref{Phi-series} cannot be resummed directly, only through its Borel transform. After the resummation of \eqref{phi-borel}, its inverse \eqref{inv-borel} should be taken to compute the function $\Phi^{(\delta)}$. However, for an asymptotic series, the function \eqref{phi-borel} produces poles and cuts on the Borel plane, and the integration contour in \eqref{inv-borel} crosses these singularities. To avoid them, one has to shift the integration contour in \eqref{inv-borel} slightly below or above the real axis and define the lateral Borel summations:
\begin{equation}\label{lateral}
    \mathcal S_\pm\left[\Phi^{(\delta)}\right](g)=g\int_{0}^{\infty e^{\pm i\epsilon}} ds e^{-gs}\mathcal{B}\left[\Phi^{(\delta)}\right](s)\,.
\end{equation}
For an asymptotic series, it produces an ambiguity in calculating the resummation of \eqref{Phi-series} and hence in computing the function $F(g)$. Due to this ambiguity, different terms in the expansion depend on the choice of regularization, and one has to choose either from:
\begin{equation}\label{Fp}
    F_+(g)=\sum_{\delta}e^{-g \sum_{l\in \delta}  x_l }g^{\lambda(\delta)}\mathcal{S}_+\left[\Phi^{(\delta)}\right](g)\,,
\end{equation}
or:
\begin{equation}\label{Fm}
    F_-(g)=\sum_{\delta}e^{-g \sum_{l\in \delta}  x_l }g^{\lambda(\delta)}\mathcal{S}_-\left[\Phi^{(\delta)}\right](g)
\end{equation}
to compute $F(g)$. We will see that due to the presence of exponentially small corrections, the ambiguities kill each other, and the resummation of $F(g)$ becomes independent of the regularization, namely: $F_+(g)=F_-(g)$.

The difference between the two prescriptions in \eqref{lateral} contains all the information about the full discontinuity of $\mathcal{B}\left[\Phi^{(\delta)}\right](s)$ along the real axis:
\begin{equation}
    \mathcal{S}_+-\mathcal{S}_-=\mathcal{S}_-\circ\text{Disc}\,.
\end{equation}
Therefore, it is convenient to introduce the so called Stokes automorphism $\mathfrak S$, which connects the two lateral resummations, and tells us how $\mathcal{B}\left[\Phi^{(\delta)}\right](s)$ jumps across the real line:
\begin{equation}\label{stokes-disc}
    \mathcal{S}_+=\mathcal{S}_-\circ\left(\mathbb I-\text{Disc}\right)\equiv \mathcal{S}_-\circ \mathfrak S\,.
\end{equation}

For the simplest example, one can take the Borel transform \eqref{phi-borel} of $\Phi^{(\delta)}$, with its large $g$ coefficients given in \eqref{phi-asymp}. Then it is straightforward to show that the function exhibits a singularity at $s=A$. The discontinouity across this singularity is:
\begin{equation}
    \lim_{\epsilon\to0}\left(\mathcal B[\Phi^{(\delta)}](s+i\epsilon)-\mathcal B[\Phi^{(\delta)}](s-i\epsilon)\right)=2i\theta(s-A)\sum_{j\geq 0}c_j\frac{(s-A)^{j-\lambda}}{\Gamma(1+j-\lambda)}\,.
\end{equation}
Therefore, the difference between the lateral Borel resummations is:
\begin{equation}\label{S-disc}
    \mathcal{S}_+[\Phi^{(\delta)}](g)-\mathcal{S}_-[\Phi^{(\delta)}](g)=2ig^{\lambda}e^{-gA}\sum_{j\geq 0}\frac{c_j}{g^j}\,.
\end{equation}
From this simple example, we immediately see that for an asymptotic series, the exponential corrections are related to the ambiguity in the resummation. Since the discontinouity is directly related to the Stokes automorphism, if one can compute $\mathfrak S$ for a transseries of the form \eqref{F-toy}, then the asymptotic expansions (namely the parameters $A$, $\lambda$ and $c_j$) of the coefficients $\phi_k^{(\delta)}$ can be determined. 

As was mentioned previously, the asymptotic series in \eqref{phi-asymp} could contain several similar terms, producing different singularities and cuts on the Borel plane. To separate the jumps along different cuts on the real line, it is useful to write the Stokes automorphism in terms of the Alien derivatives. By definition, they are related to $\mathfrak S$ via:
\begin{equation}\label{Stokes-aut}
    \mathfrak S =\exp\left(\sum_{\omega}e^{-\omega g}\Delta_\omega\right)\,,
\end{equation}
where the summation runs along all the singular points $\omega$ of the Borel transform. $\Delta_\omega$ called a derivative, since it satisfies the Leibnitz rule.

The calculation of the Alien derivatives can be done via the so called Bridge equations. The first step is to assign a parameter $\sigma_j$ to each exponential weight $x_j$ in the transseries \eqref{F-toy} and define:
\begin{equation}
    F(g,\{\sigma\})=\sum_{\delta}\left(\prod_{l\in\delta}\sigma_l\right) e^{-g \sum_{l\in \delta}  x_l }g^{\lambda(\delta)}\Phi^{(\delta)}(g)\,,
\end{equation}
so every exponential factor $e^{-gx_l}$ comes with a factor of $\sigma_l$. Now it can be shown (see \citep{Marino:2012zq,Dorigoni:2014hea,Aniceto:2018bis}), that the combination:
\begin{equation}\label{dotdef}
    \dot\Delta_\omega=e^{-\omega g}\Delta_\omega
\end{equation}
commutes with differentiation with respect to the variable $g$. The same holds for differentiation with respect to the parameters $\sigma_j$, so that:
\begin{equation}
    \left[\partial_g,\partial_{\sigma_j}\right]=\left[\partial_g,\dot{\Delta}_\omega\right]=0\,.
\end{equation}
If we suppose that $F(g)$ is a solution of any non-linear problem, it follows that $\partial_{\sigma_j}F(g,\{\sigma\})$ (with every parameter $\sigma_j$) and $\dot{\Delta}_\omega F(g,\{\sigma\})$ solve the same linear, homogeneous equation whose exact form depends on the original problem. As a consequence, $\dot\Delta_\omega$ can be written as a linear combination of differentiations with respect to the parameters $\sigma_j$:
\begin{equation}\label{dot-prop}
    \dot\Delta_\omega=\sum_{l\geq0}A_{\omega,l}\partial_{\sigma_l}\,,
\end{equation}
where $A_{\omega,l}$ are related to the Stoke constants.
This is called the Bridge equation and it relates the partial derivatives with respect to the resurgence parameters $\sigma_j$ and the Alien derivatives $\Delta_j$.

Acting on $F(g,\{\sigma\})$ with \eqref{dotdef} and the right-hand side of \eqref{dot-prop}, and comparing the exponential factors and different powers of the parameters $\sigma_j$, one finds that $\omega$ should be one of the exponential weights $x_l$. Furthermore, the different non-perturbative sectors are related as:
\begin{align}\label{non-zeroAD}
    &\Delta_j\Phi^{(\delta)}=g^{\lambda(\delta\cup\{j\})-\lambda(\delta)}A_j\Phi^{(\delta\cup\{j\})}\,,
\end{align}
where $j\notin\delta$, and for simplicity, I introduced the notation $\Delta_{x_j}\equiv \Delta_j$ and relabeled the coefficients $A^\pm_{x^\pm_j,j}\equiv A^\pm_{j}$. It also follows from the Bridge equations that:
\begin{equation}\label{Alien-algebra}
    \left(\Delta_j\right)^2=0,\qquad \left[\Delta_j,\Delta_l\right]=0\,,
\end{equation}
for every integer $j,l\geq 0$. 

With the help of the algebra in \eqref{Alien-algebra}, the exponentiation in \eqref{Stokes-aut} can easily be evaluated to get a more reasonable expression for the Stokes automorphism. From \eqref{Alien-algebra} it immediately follows that $\mathfrak S$ is formally equal to:
\begin{align}\label{stokes-simp}
    \mathfrak S&=\prod_{l\geq 0}\left(1+e^{-x_l g}\Delta_l\right)=1+\sum_{j\geq1}\sum_{0\leq n_1\leq\dots n_j}e^{-g(x_{n_1}+x_{n_2}+\dots +x_{n_j})}\Delta_{n_1}\Delta_{n_2}\dots\Delta_{n_j}\,.
\end{align}

Acting on the function $\Phi^{(\delta)}$ with \eqref{stokes-simp} and using the definition of $\mathfrak S$ from \eqref{stokes-disc}, one finds for the lateral Borel resummations $\mathcal{S}_\pm\left[\Phi^{(\delta)}\right](g)$ that:
\begin{align}\label{gen-disc}
    \mathcal{S}_+\, \Phi^{(\delta)}=\mathcal{S}_-\Phi^{(\delta)}+\sum_{j\geq1}\sum_{\substack{0\leq n_1\leq\dots n_j\\\forall l\leq j,\,n_l\notin\delta}}g^{\lambda(\delta\cup\{n_1,\dots,n_j\})-\lambda(\delta)}e^{-g(x_{n_1}+x_{n_2}+\dots +x_{n_j})}A_{n_1}A_{n_2}\dots A_{n_j}\mathcal{S}_-\Phi^{(\delta\cup\{n_1,\dots,n_j\})}\,.
\end{align}
This relation means, that the Borel transform of $\Phi^{(\delta)}$ contains cuts starting from every point $x_{n_1}+x_{n_2}+\dots +x_{n_j}$ on the real line, with each $x_{n_l}\notin \delta$. According to \eqref{non-zeroAD}, $\Phi^{(\delta)}$ has direct resurgence relations only with the exponential corrections $\Phi^{(\delta\cup\{j\})}$ with $j\notin \delta$. The rest of the terms in \eqref{gen-disc} (and hence in the asymptotic expansion of $\Phi^{(\delta)}$) come from the asymptotic behavior of higher order corrections $\Phi^{(\delta\cup\{j\})}$. 

By substituting \eqref{gen-disc} into $F_+(g)$ in \eqref{Fp}, higher order terms in \eqref{gen-disc} kill each other and the series becomes equal to $F_-(g)$ in \eqref{Fm}. This means that the ambiguities cancel and $F(g)$ becomes independent of the regularization: $F(g)=F_+(g)=F_-(g)$. 

At first, the expression in \eqref{gen-disc} looks tedious; however, taking specific values of $\delta$ clarifies that \eqref{gen-disc} is enough to determine the asymptotic behavior of the functions $\Phi^{(\delta)}(g)$ and a possible way to compute the constants $A_j$ that appear in the Alien derivatives in \eqref{non-zeroAD}. Consider, for example, the set $\delta=\{0,1,\dots,l\}$ that contains all integers from $0$ to $l$. For the corresponding functions $\Phi^{(\{0,1,\dots,l\})}(g)$, the expression in \eqref{gen-disc} looks as:
\begin{equation}\label{simp-disc}
    \mathcal{S}_+\Phi^{(\{0,\dots,l\})}-\mathcal{S}_-\Phi^{(\{0,\dots,l\})}=g^{\lambda(\{0,\dots,l+1\})-\lambda(\{0,\dots,l\})}e^{-gx_{l+1}}A_{l+1}\mathcal{S}_-\Phi^{(\{0,\dots,l+1\})}+\dots\,,
\end{equation}
where the dots denote exponentially suppressed terms. Now suppose that $\Phi^{(\{0,\dots,l\})}(g)$ is given by a series in $1/g$ as in \eqref{Phi-series} and its coefficients $\phi^{(\{0,\dots,l\})}_k$ can be parametrized as:
\begin{equation}
    \phi_k^{(\{0,\dots,l\})}=\frac{1}{\pi}\sum_{j\geq 0}c_j\frac{\Gamma(k+\lambda-j)}{A^{k+\lambda-j}}+\dots\,,
\end{equation}
with the dots standing for higher order cuts. Then by \eqref{S-disc} and \eqref{simp-disc} it follows that the parameters $A$, $\lambda$ and $c_j$ in the asymptotic series are given as:
\begin{align}\notag\label{asym-params}
    A&= x_l\,,\notag\\
    \lambda&=\lambda(\{0,\dots,l+1\})-\lambda(\{0,\dots,l\})\,,\notag\\
    c_j&= \frac{A_{l+1}}{2i}\phi^{(\{0,\dots,l+1\})}_k\,,
\end{align}
where $\phi^{(\{0,\dots,l+1\})}_k$ are the $1/g$ coefficients of $\Phi^{(\{0,\dots,l+1\})}(g)$. Therefore, the leading asymptotic expansion of $\phi_k^{(\{0,\dots,l\})}$ is:
\begin{equation}\label{phi-k-0-l}
    \phi_k^{(\{0,\dots,l\})}=\frac{A_{l+1}}{2\pi i}\sum_{j\geq 0}\phi^{(0,\dots,l+1)}_j\frac{\Gamma(k+\lambda-j)}{(x_l)^{k+\lambda-j}}+\dots\,,
\end{equation}
with $\lambda$ given in \eqref{asym-params}. This means that by investigating the leading asymptotic behavior of $\phi_k^{(\{0,\dots,l\})}$ (for example, with the method discussed in Section 7 and Appendix G of \cite{Bajnok20252}), all coefficients $A_{l+1}$ can be systematically determined.

\bibliographystyle{JHEP}    
\bibliography{references}      
     
\end{document}